\documentstyle[12pt,epsf]{article}

\begin{document}

\newcommand{\beq}{\begin{equation}}
\newcommand{\eeq}{\end{equation}}
\newcommand{\bzp}{\tilde{\beta}_+}
\newcommand{\bzm}{\tilde{\beta}_-}
\newcommand{\bzpc}{\tilde{\beta}^{2}_+}
\newcommand{\bzmc}{\tilde{\beta}^{2}_-}
\newcommand{\bzpt}{$\tilde{\beta}_+$}
\newcommand{\bzmt}{$\tilde{\beta}_-$}
\def\z3g{Z \rightarrow 3\gamma}
\def\z3gt{$Z \rightarrow 3\gamma\;$}

\newcommand{\beqn}{\begin{eqnarray}}
\newcommand{\eeqn}{\end{eqnarray}}

\newcommand{\su}{$ SU(2) \times U(1)\,$}

\newcommand{\np}{Nucl.\,Phys.\,}
\newcommand{\pl}{Phys.\,Lett.\,}
\newcommand{\pr}{Phys.\,Rev.\,}
\newcommand{\prl}{Phys.\,Rev.\,Lett.\,}
\newcommand{\prep}{Phys.\,Rep.\,}
\newcommand{\zp}{Z.\,Phys.\,}

\newcommand{\mw}{M_{W}}
\newcommand{\mww}{M_{W}^{2}}
\newcommand{\mz}{M_{Z}}
\newcommand{\mzz}{M_{Z}^{2}}
\newcommand{\ie}{{\em i.e.}}
\newcommand{\epm}{$e^{+} e^{-}\;$}
\newcommand{\eps}{\varepsilon}
\def\epem{\ifmmode{{e^{+}e^{-}}}\else{${e^{+}e^{-}}$}\fi}
\newcommand{\ra}{\rightarrow}
\newcommand{\lra}{\leftrightarrow}
\newcommand{\tr}{{\rm Tr}}

\newcommand{\gag}{$\gamma \gamma$ }

\newcommand{\dkg}{\Delta \kappa_{\gamma}}
\newcommand{\dkz}{\Delta \kappa_{Z}}
\newcommand{\dz}{\delta_{Z}}
\newcommand{\dgz}{\Delta g^{1}_{Z}}
\newcommand{\la}{\lambda}
\newcommand{\lag}{\lambda_{\gamma}}
\newcommand{\laz}{\lambda_{Z}}
\newcommand{\lnl}{L_{9L}}
\newcommand{\lnr}{L_{9R}}
\newcommand{\lt}{L_{10}}
\newcommand{\lu}{L_{1}}
\newcommand{\ld}{L_{2}}

\def\bm{\beta_-}
\def\bp{\beta_+}
\def\hm{{\cal H}_-}
\def\hp{{\cal H}_+}
\def\hpm{{\cal H}_\pm}
\def\calf{{\cal F}}
\def\calg{{\cal G}}
\def\cald{{\cal D}}
\def\call{{\cal L}}
\def\bremplus{\sin^2\frac{\tau\hp^{1/2}}{2}}
\def\sinminus{\sin^2\frac{\tau\hm^{1/2}}{2}}
\def\cosmp{[\cos(\tau\hm^{1/2})-\cos(\tau\hp^{1/2})]}

\newcommand{\cm}{{{\cal M}}}
\newcommand{\cl}{{{\cal L}}}
\newcommand{\cd}{{{\cal D}}}
\newcommand{\cv}{{{\cal V}}}
\def\slashc{c\kern -.400em {/}}
\def\slashL{L\kern -.450em {/}}
\def\slashcl{\cl\kern -.600em {/}}
\def\W{{\bf W}}
\def\B{{\bf B}}
\def\noi{\noindent}
\def\sm{${\cal{S}} {\cal{M}}\;$}
\def\nph{${\cal{N}} {\cal{P}}\;$}
\def\ssb{${\cal{S}} {\cal{S}}  {\cal{B}}\;$}
\def\cviol{${\cal{C}}\;$}
\def\pviol{${\cal{P}}\;$}
\def\cpviol{${\cal{C}} {\cal{P}}\;$}

\begin{titlepage}
\begin{center}

{ \Large \bf New Physics with three-photon events at LEP}

\vspace*{1.cm}

\begin{tabular}[t]{c}

{\bf M.~Baillargeon$^{1}$,  F.~Boudjema$^{2}$, E. Chopin$^{2}$ and V.
Lafage$^{2}$ }\\
\\
{\it 1.  Grupo Te\'orico de Altas Energias, Instituto Superior T\'ecnico}\\
{\it Edif\'{\i}cio Ci\^encia (F\'{\i}sica)
P-1699 Lisboa Codex, Portugal}\\
%
%
\\
%
{\it 2. Laboratoire de Physique Th\'eorique}
EN{\large S}{\Large L}{\large A}PP
\footnote{URA 14-36 du CNRS, associ\'ee \`a l'E.N.S de Lyon et \`a
l'Universit\'e de Savoie.}\\
{\it Chemin de Bellevue, B.P. 110, F-74941 Annecy-le-Vieux, Cedex, France.}

\end{tabular}
\end{center}
\vspace*{\fill}

\centerline{ {\bf Abstract} }
\baselineskip=14pt
\noindent
 {\small  The effect of the most general \z3gt vertex in the reaction
$e^+e^- \ra 3\gamma$ is studied with a particular attention to LEP searches.
We give exact analytical expressions including realistic cuts for the signal
and
present a detailed analysis based on a Monte Carlo that includes the effect
of the irreducible $3\gamma$ QED cross section. As special applications
we discuss the effect of heavy scalars, fermions and gauge bosons and comment
 on the ``monopole" connection.}
\vspace*{\fill}


\vspace*{0.1cm}
\rightline{ENSLAPP-A-518/95}
\rightline{FISIST/6-95/CFIF}
\rightline{hep-ph/9506396}
\rightline{May 1995}
\vspace*{\fill}

\noindent
{\footnotesize $\;^{\S}$ URA 14-36 du CNRS, associ\'ee \`a l'E.N.S de Lyon et
\`a
l'Universit\'e de Savoie.}

\end{titlepage}
\baselineskip=18pt

\setcounter{section}{1}

\setcounter{subsection}{0}
\setcounter{equation}{0}
\def\thesubsection {\thesection.\arabic{subsection}}
\def\theequation{\thesection.\arabic{equation}}

\setcounter{equation}{0}
\def\thequation{\thesection.\arabic{equation}}

\setcounter{section}{0}
\setcounter{subsection}{0}

\section{Introduction}
The second phase of LEP will soon be in operation. The achievements of
the first phase have already been formidable as we have learnt a great deal
from the precision measurements at the $Z$ peak on the parameters of the
Standard Model, \sm. Some of the future analyses will need to exploit   the
 full statistics to be accumulated
at the end of the LEP1 runs\footnote{By the end of 1994 each LEP
collaboration  has collected about 140~pb$^{-1}$
($\sim 5.5 \; 10^6 Z$) of data samples but only a fraction of these has
been used in published analyses of rare processes. At the end of the LEP1
runs one hopes to accumulate about 200~pb$^{-1}$ per experiment
\cite{Maettig}.}
 in order to put stringent constraints on rare processes
that are usually the hallmark of New Physics. One such process is
the decay of the $Z$ into three photons. Within the \sm this process is so rare
that even the once-discussed High-Luminosity option of LEP1   \cite{Highlumi}
would not be able to
see it. Indeed, branching ratio for this process calculated within the \sm is
$Br_{sm}(Z\ra 3\gamma)\simeq 5.4\; 10^{-10}$   \cite{Z3gammasm}.  This makes
it an
ideal candidate to search for contributions that are beyond the \sm. It is
interesting to note that this kind of quartic vertex, $Z3\gamma$, is an example
of an
``anomalous" boson self-coupling which can be probed at LEP1 whereas the second
phase of LEP will be mostly dedicated to the tri-linear $WW\gamma,WWZ$
couplings. The study of this coupling thus provides a nice bridge, when moving
from
 LEP1 to LEP2,  in the general topic
of the {\em direct}\footnote{Of course, LEP1 is of an unsurpassable
precision when it comes to the investigation of the vector bosons self-energies
and hence on what might be considered as bi-linear anomalous couplings.}
 investigations on the vector boson self-couplings. From the experimental point
of view, another  motivation for studying this decay is the observation that
the
description of this rare process is, as we shall see,  identical to the
description of the scattering of light by light parameterised by a 4-$\gamma$
coupling. Although QED has been with us for a long time, this non-linear
effect has not been directly experimentally investigated. As far as we are
aware,
one has only experimentally studied Delbr\"uck scattering and photon splitting
near a heavy nucleus at low energy   \cite{Delbruck}. One should therefore not
miss
any opportunity of investigating a very similar kind of physics, especially
since
the theoretical situation at LEP1 with the \z3gt is much cleaner.  Thus one
should
take full advantage of the resonance enhancement at the $Z$ peak which
provides an alternative to study New Physics contributions to  an  Abelian
non-linear effect.

Before the advent of LEP, it had been argued   \cite{Fernand3g,YellowbookLep1}
that this decay could check the
hypothesis of a composite $Z$. The argument heavily borrows from the $\rho$
system and vector meson dominance   \cite{Sakurai}. To explain the universality
of the weak
coupling  in this picture, one adapts the vector dominance idea to the
$W$ triplet by implementing a $W^0-\gamma$ transition   \cite{HungSakurai}. The
universality of the
$QED$ electromagnetic coupling is thus transmitted to the $W/Z$-fermion
coupling. The difference between the $\rho$ system and the $W$ is that the
strength of this transition, directly related to the weak mixing angle $s_W$,
is not
small as would be expected for an electromagnetic transition but of order
$1/2$.
Since this is large,  one should expect that other electromagnetic
transitions be enhanced   \cite{Normanzzg}. More specifically, in this picture,
 one can build up on the
``strength" of the $Z\gamma$ transition to predict a large decay of the $Z$
into
three photons. Of course, the $Z$ can not decay into two on-shell  photons
(Yang's
theorem   \cite{Yang})
and in any case the $Z2\gamma$ vertex violates \cviol and \pviol and thus could
be
forbidden. However, \z3gt is perfectly allowed. Arguing along these lines, a
{\em non-relativistic} bound state
calculation   \cite{Fernand3g} of the \z3gt
leads to a branching ratio of the order of  $10^{-5}$. This
is in fact  the present order of magnitude the LEP
experiments   \cite{Aleph,Delphi,L3,Opal} have set, as  a limit, on
this decay.

Recently De R\'ujula   \cite{DeRujula3g} has suggested that this decay may be
large
if it is induced by ``monopoles". This idea has already been put
forth   \cite{Ginzburg4g} a few
years ago to motivate a non-negligible 4-$\gamma$ coupling. The aim of this
paper is, first, to give the most general framework for the study of the $Z \ra
3\gamma$ where the above two examples ( so-called ``composite" $Z$ and
``monopoles")
will be seen to be specific cases of the
most general effective Lagrangian. We will then give a detailed analysis on the
signature of the $Z\ra 3\gamma$ decays. For the most general Lagrangian
we  give exact analytical formulae  for various distributions including the
effect of the correlation with the initial \epem
state. We also derive analytical formulae
taking into account experimental cuts that can be used for a quick estimate of
the acceptances of the 4 LEP detectors. Finally we conduct a detailed
discussion  based on a Monte Carlo analysis which includes not only the signal
but also the irreducible $\epem \ra 3\gamma$ purely QED background as well as
the effect of the interference. We show which distributions are most sensitive
to the presence of the $Z\ra 3\gamma$ coupling. Finally a short discussion
about
the effect of this coupling far away from the $Z$ resonance is presented. In
the
Appendix we collect some simple and compact expressions for the standard and
non-standard  helicity amplitudes based on the technique of the inner spinor
product
that were implemented in our matrix element Monte Carlo generator.

\setcounter{subsection}{0}
\setcounter{equation}{0}
\def\thesubsection {\thesection.\arabic{subsection}}
\def\theequation{\thesection.\arabic{equation}}
\def\thefigure{\arabic{figure}}

\section{Effective Lagrangians and Models for \boldmath{\z3gt}}
Probably the oldest example of an effective Lagrangian is the celebrated Euler
Lagrangian   \cite{Euler} that describes the self-interaction of photons.
 The first non-trivial
interaction describes a 4-$\gamma$ vertex in its leading part
and accounts for the scattering of light-by-light. The basic idea of the
effective
Lagrangian is to give a general description of a phenomenon even if one does
not know its origin or the underlying theory behind it. All we need to
construct
the ensuing operators is the known symmetries at low energies which are, in
this example,
contained in the $U(1)$
gauge invariance. For the 4-$\gamma$ vertex one is led to
\beqn
\label{Lag4g}
\cl_{eff.}^{4\gamma}=  \frac{\alpha^2}{M^4}  \left(
\beta_2 (F_{\mu \nu} F^{\mu \nu})^2
\;+\;\beta_3 (F_{\mu \nu} \tilde{F}^{\mu \nu})^2 \right)
\;\;\;\;\;\;\hbox{with} \;\;\;\;\;
\tilde{F}^{\mu \nu}=\frac{1}{2}
\varepsilon^{\mu \nu \alpha \beta} F_{\alpha \beta}.
\eeqn
We remark that even though one is dealing with an effective Lagrangian,
and restricting oneself to the {\em leading} operators, the Lagrangian is
very constrained  since it contains only two operators. These are necessarily
of
dimension eight. The mass $M$ is directly related to the scale of New Physics.
Non-leading operators contain extra derivatives that give
correction factors of order $(E_\gamma/M)^2$ to the dominant terms we are
considering, and therefore their effect is very negligible. We can also argue
that if these terms were  to be taken into account then this would {\em also}
mean that
multi-photon amplitudes $6\gamma, ....$ etc would not be negligible and thus
the
New Physics would be more conspicuous.

Different models, or rather types of heavy particles, give definite predictions
for $\beta_{2,3}$. For instance, considering the case of a massive particle
(mass
$M$) of unit charge that contributes to the one-loop 4-$\gamma$ amplitude, we
have in
the case of fermions   \cite{Euler,Schwinger}
\beq
\label{floop4g}
\beta_2^F=\frac{4}{7} \beta_3^F=\frac{1}{90}.
\eeq
For a charged scalar the coefficients $\beta_i$'s are different with
  \cite{Schwinger}
\beq
\label{sloop4g}
\beta_2^S=7 \beta_3^S=\frac{7}{16} \frac{1}{90}.
\eeq
For completeness, we can also give the contribution of a heavy charged
gauge vector boson. Here as the spin increases the values of the coefficients
increase as well.  This is not only due to the counting of degrees
of freedom.
Although the coefficients for the $\beta's$ in the case of scalars and fermions
have been known for a long time   \cite{Schwinger}, it is only in the last few
years that the issue
of the vector, more specifically the $W$-loop, has been investigated
\cite{Wloop4g}. The
 $\beta's$ can be inferred from   \cite{Wloop4g}:
\beq
\label{coefwloop4g}
\beta_2^V=\frac{29}{160} \;\;\;\;\; \beta_3^V=\frac{27}{160}.
\eeq
It is instructive to realise that if instead of considering $\beta_{2,3}$ one
takes the combination
\beq
\beta_\pm=\beta_{2}\pm \beta_{3}
\eeq
the counting of the degrees of freedom, with the correct Bose-Fermi
statistics factor, is reflected only in $\beta_-$:
\beqn
\beta_+^S=\frac{1}{180}  \;\;\;&;&\;\;\; \beta_-^S =\frac{1}{240}
\nonumber \\
\beta_+^F=\frac{11}{360} \;\;\;&;&\;\;\;
\beta_-^F=-2\;\beta_-^S \\
\nonumber \\
\beta_+^V=\frac{7}{20} \;\;\;&;&\;\;\; \beta_-^V=3\;\beta_-^S. \nonumber
\eeqn
We will have more to say about the extraction of these coefficients and
about the so-called supersymmetric relations   \cite{Susyrelations} between the
$\beta_-$
elsewhere   \cite{Usdeterminant}.

The $Z\ra 3\gamma$ vertex is constructed along the same
mould. In fact, since we are replacing {\em only one} photon by a $Z$ the
general
leading-order effective Lagrangian has exactly the same structure. Therefore,
we
choose to parameterise the $Z3\gamma$ anomalous coupling in the form
\beqn
\label{Lagz3g}
\call_{Z3\gamma}&=& \frac{4\alpha^2}{M^4} \left(
\tilde{\beta}_2 (F_{\mu \nu} F^{\mu \nu}) (F_{\mu \nu} Z^{\mu \nu})
\;+\;\tilde{\beta}_3 (Z_{\mu \nu} \tilde{F}^{\mu \nu})
(F_{\mu \nu} \tilde{F}^{\mu \nu})  \right) \nonumber \\
& \equiv &
\frac{4\alpha^2}{M^4} \left(
(\tilde{\beta}_2-2 \tilde{\beta}_3)(F_{\mu \nu} F^{\mu \nu}) (F_{\mu \nu}
Z^{\mu \nu})
\;+\;\ 4 \tilde{\beta}_3 (Z_{\mu \nu} F^{\nu \lambda} F_{\lambda \rho} F^{\rho
\mu})
\right).
\eeqn
The factor of 4 in the definition of the $Z3\gamma$ coupling compared to the
case of the $4\gamma$ coupling is such that the two Lagrangians lead to the
same
strength, and expression in momentum space, for the $Z3\gamma$ and $4\gamma$
coupling
modulo the overall change
$\beta_i \ra \tilde{\beta}_i$.

We refrain here from giving specific examples of  the coefficient
$\tilde{\beta}_{2,3}$
for the different spin species as, beside the electric charge of the particles,
we need to know their hypercharge or $SU(2)_{weak}$ assignment. Nonetheless
the relations between $\tilde{\beta}_2$ and $\tilde{\beta}_3$ are maintained in
the same
form as in the all 4-$\gamma$ case (Eq.~\ref{floop4g}-~\ref{sloop4g}),
apart from the case of
the spin-1, where there is a subtlety. For the latter one needs to know both
the quantum number of the pure gauge part {\em and} the Goldstone part
of the massive vector bosons. In practice the pure gauge part is by far
dominant. Since these are rather technical observations and calculations we
leave this discussion to a forthcoming paper   \cite{Usdeterminant}.

Again it is worth working, instead of the two independent constants
$\tilde{\beta}_2$ and
$\tilde{\beta}_3$, with  the combination
\beq
\tilde{\beta}_\pm=\tilde{\beta}_2 \pm \tilde{\beta}_3.
\eeq

The reason is that $\tilde{\beta}_+$ and $\tilde{\beta}_-$ contribute to
different
helicity amplitudes and therefore they do not interfere.
It is easy to see the reason why this is
so. For simplicity it is best to stick with the 4-$\gamma$ Lagrangian and
rewrite it  in terms of the self-dual, $F^+_{\mu \nu}$, and anti-self-dual,
$F^-_{\mu \nu}$, fields which for photons correspond to
{\em definite} helicity
states, namely
\beq
F_{\pm}^{\mu \nu}=\frac{1}{2} (F^{\mu \nu}\pm i \;\tilde{F}^{\mu
\nu})\;\;\;\; ; \;\;\; F_-\cdot\;F_+=F_{-}^{\mu \nu}\ F_{+\;\mu \nu}=0.
\eeq
In this case one has
\beqn
\cl_{eff.}^{4\gamma}\equiv\frac{\alpha^2}{M^4}\left(
2\beta_+\; F_+^2 F_-^2 \;\;+\;\;\beta_- \; \left( ( F_+^2)^2 \;+\; ( F_-^2)^2
\right)\right).
\eeqn
It is then obvious that the terms in $\beta_+$ contribute to different
helicity amplitudes than those with $\beta_-$:  formally taking a configuration
 with only either a self-dual or an
anti-self-dual (either $F^+ \ra 0$ or $F^-\ra 0$) the $\beta_+$ does not
contribute. For the same reason, in the process \z3gt based on the identical
Lagrangian, only
$\bzm$ contributes to the amplitude where all $3\gamma$
have the same helicity.

For the discussion we will consider 5 typical one-parameter ``models" of
$\beta$:
\begin{itemize}
\item $\tilde{\beta}_{2,3}$: As stressed above, since it is only the
combination
$\tilde{\beta}_\pm$ that appears at the helicity amplitude level, it is not
possible
to disentangle a model with
only $\tilde{\beta}_2$ from a model with only $\tilde{\beta}_3$ in three-photon
production.
The model with $\tilde{\beta}_2$ has, in an abuse of terms, been referred to as
a ``composite" $Z$.
$\tilde{\beta}_2$ could, more appropriately, be associated with the effect of a
heavy scalar having a
two-$\gamma$ and $Z\gamma$ coupling, while $\tilde{\beta}_3$ with that of a
pseudo-scalar.  Two of
us   \cite{Yellowbookfinalg} have devoted a preliminary analysis to the effect
of the  operator
$\tilde{\beta}_2$
some time ago.
\item a model with $\tilde{\beta}_+$ but $\tilde{\beta}_-=0$
\item a model with $\tilde{\beta}_-$ but $\tilde{\beta}_+=0$
\item a model that we will refer to as the $\beta_4$ model
with $\tilde{\beta}_+=-11/3\; \tilde{\beta}_-$. This is the same
model as that considered in   \cite{DeRujula3g}. It corresponds to the effect
of
a heavy fermion which in   \cite{DeRujula3g} is considered to be a monopole.
In  \cite{DeRujula3g} the exotic fermion is assumed to be a $SU(2)$ singlet.
With this information it is straightforward to relate the $4\gamma$ and
$Z3\gamma$
strengths as
\beq
\label{fmonopole}
\tilde{\beta}_{2,3}=\frac{s_W}{c_W} \beta_{2,3}^F \;\;\;\;\;
\;\; (c_W^2=1-s_W^2)
\eeq
where $\beta_{2,3}^F$ are given by Eq.~\ref{floop4g}. Taking the fermion to be
a point-like monopole (free isolated magnetic charge) one is also led
to assume that the strength of its coupling to the photon, $g_m$, is strong. In
fact Dirac's quantification   \cite{Dirac} condition relates this coupling to
the electric coupling through
\beq
e g_m=2\pi n\;\;\;\; (n=\hbox{ integer}).
\eeq
Therefore in our parameterisation one has to allow for the change $e\ra g_m$ in
Eq.~\ref{Lagz3g}. Although one may be tempted by giving a monopole motivation
to this type of coupling, it remains that one still has two parameters:
the mass of the monopole and the magnetic strength, $g_m$
(or quantification number $n$) which are lumped in one effective parameter
$M/n$,
exactly as what one has
with the effective Lagrangian (Eq.~\ref{Lagz3g}). The discriminating relation
Eq.~\ref{fmonopole} indicates the spin-$1/2$ nature of the point-like
monopole.
Note, for the record, that the idea of a monopole to induce a large 4-$\gamma$
coupling
is not new; it has been exploited some time ago by Ginzburg and Panfil
\cite{Ginzburg4g}.
They also considered the case of a spin-0 monopole which again can be
accommodated by the general effective Lagrangian through the discriminating
relation $\tilde{\beta}_{2,3}=\frac{s_W}{c_W} \beta_{2,3}^S$ (and $e \ra g_m$);
however, from what we
remarked earlier, for the same quantification condition the spin-$1/2$ monopole
gives a larger contribution.

\item In fact, by far, the most interesting scenario is that of
the effect of a  heavy spin-1 boson. This could be a model with
$\tilde{\beta}_+=28 \tilde{\beta}_-$.
This corresponds
to the effect of a very heavy vector boson whose pure gauge component and
Goldstone
component have the same quantum numbers. This may be considered as the vector
boson version of the ``monopole" considered by De R\'ujula   \cite{DeRujula3g}.
In practice this model is very similar
to a model with $\tilde{\beta}_-=0$ in view of the much larger value of
$\tilde{\beta}_+$, especially if one keeps  in mind  that the main
contributions of these operators,
as we shall see, are quadratic
in the couplings.
For a thorough discussion on these models and others
we refer to  \cite{Usdeterminant}.
\end{itemize}

In the monopole vein, one should not dismiss the eventuality that due to the
necessarily very strong coupling of the monopoles, bound states may form with a
mass that is much smaller than their combined constituent masses. One
interesting possibility would be that the ground state is of spin-0 that decays
predominantly into $\gamma \gamma$ and $Z\gamma$, therefore dynamically
realising models with $\tilde{\beta}_{2,3}$. In the remaining of the paper we
will
steer away from considerations pertaining to the manifestation of the
monopoles.
This is
not only because one would have to argue that their masses are, against what is
generally but not universally {\em believed}, much  below the
GUT scale (in order to evade a Lilliputian  unobservable effective \z3gt
coupling),
but also because one has to quantify their effect on the two-point functions
that
are extremely well measured at LEP1 and contrast the information with the one
given by \z3gt\footnote{Moreover, beside the controversial
``lightness" ($\sim$~TeV) and point-like nature of the Dirac monopole,
consistency of QED in the presence of a Dirac monopole has been
questioned   \cite{Monopolecontroversy}.}.
We will stick to the general unbiased approach of studying the
manifestation of a general \z3gt coupling referring to the models as defined
through the relation between $\bzp$ and $\bzm$. We take the view that the
``monopole" is just a  possible  paradigm for the existence of a not too small
\z3gt coupling.

We also note for further discussion that the authors
of reference  \cite{Tcheques} have also given a parameterisation in terms
of two operators but have only studied the double integrated \z3gt partial
width
and the totally integrated width. To make contact with their parameters we note
that their constants
$G_{1,2}$ are related to ours as
\beq
G_1=\frac{32 \alpha^2}{M^4} (\tilde{\beta}_3-\tilde{\beta}_2)=-\frac{32
\alpha^2}{M^4}\bzm \;\;\;\;\;\;\;\;
G_2=\frac{32 \alpha^2}{M^4} \tilde{\beta}_3.
\eeq

\setcounter{subsection}{0}
\setcounter{equation}{0}
\def\thesubsection {\thesection.\arabic{subsection}}
\def\theequation{\thesection.\arabic{equation}}
\def\thefigure{\arabic{figure}}

\section{Analytical Formulae for $\Gamma(Z\ra 3\gamma)$ and
$\epem\stackrel{Z}{\ra} 3\gamma$ }

{}From the general Lagrangian, it is an easy task to compute the $Z$ width into
three photons as well as the correlation with the initial state through the
cross
section $\epem \ra 3\gamma$. We will start by providing analytical
formulae for the partial width and will give compact analytical formulae
including
realistic experimental cuts, that could be considered as canonical cuts for
this process. The latter can be used to quickly estimate the acceptances.
\begin{figure*}[htb]
\begin{center}
\caption{\label{Feynman}{\em
The t-channel QED process and the s-channel anomalous \z3gt
contribution to three-photon production.}}
\mbox{\epsfxsize=11cm\epsfysize=4cm\epsffile{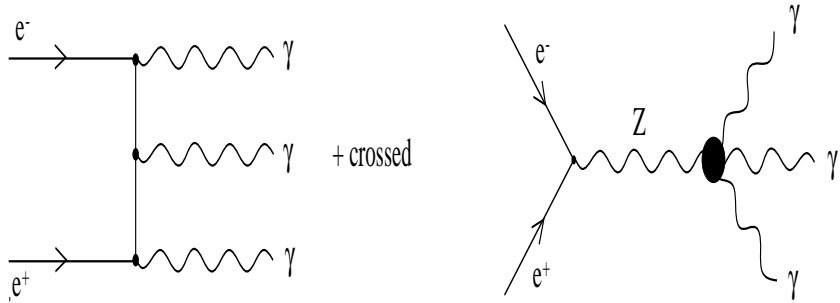}}
\end{center}
\end{figure*}

{}From the start, we would like to point out, before exhibiting any formula,
that
since the effective operator is of
dimension 8, constructed out of field strengths, the $s$-channel produced
photons, see
Fig.~1, will be hard
photons that will not be collinear to the beam. This is in contrast to the
$t$-channel QED background (see Fig.~1).  This is  a general characteristics
of photons that probe New Physics (See   \cite{Yellowbookfinalg}). In the case
at hand,  we should therefore
expect that the
distribution in the least energetic photon, for instance, will be more
sensitive
to the presence of the \z3gt vertex.

Specialising to the decay $Z \ra 3\gamma$ and taking as variables
the reduced photon energies $x_i=2 E_i/M_Z$, the double
distribution in the energies writes:
\beqn
& &\frac{1}{\Gamma}\frac{{\rm d}\Gamma}{{\rm d}x_1 {\rm d}x_2}\nonumber \\
&=&
\frac{120}{3\bzpc + 5 \bzmc}\; \left\{
(\bzpc + \bzmc) (1-x_1)(1-x_2)(1-x_3) \;\;\;+ \;\;\;
\hspace*{1.5cm} \right. \nonumber \\
&&\left.
(\bzpc +2 \bzmc) \left( (1-x_1)^2(1-x_2)^2 + (1-x_1)^2(1-x_3)^2 +
(1-x_2)^2(1-x_3)^2 \right) \right\} \nonumber \\
&\equiv&
\frac{60}{3\bzpc + 5 \bzmc}   \left\{ (\bzpc+2 \bzmc)
\left( (1-x_1)^2 x_1^2 + (1-x_1)^2 x_2^2+
((1-x_3)^2 x_3^2\right)\right.  \nonumber \\
&-& \left. 2\bzmc (1-x_1)(1-x_2)(1-x_3) \right\},
\eeqn
where the totally integrated partial width is calculated to be

\beqn
\label{width3g}
\Gamma=\frac{M_Z}{18} \frac{1}{\pi^3}
\left( \alpha \frac{M_Z^2}{M^2} \right)^4
\left( \frac{3\bzpc + 5\bzmc }{120} \right).
\eeqn

We disagree with the expression given in  \cite{DeRujula3g} in the case of
the fermion monopole, although we confirm the expression for the normalised
differential cross
section. We find a rate that is 12 times smaller than the one given
in  \cite{DeRujula3g}
\footnote{ This would mean that the extracted L3  \cite{L3} limit on the mass
of the
monopole should be corrected by a factor $\simeq 2$. Moreover from the
MC that we will provide it will be possible to implement the correct
acceptance factor for this model.}
. On the other hand, we confirm the
direct one-loop calculation of the supersymmetric charged Higgses given
in  \cite{Koenig}. Needless to say that for this weakly interacting
heavy particle the effect is hopelessly beyond observability.
We also agree with the expression given in  \cite{Tcheques}
in terms of $G_{1,2}$.
We note, in connection with the monopole case, that had we considered a spin-1
rather than
spin-1/2 exotic, the spin-1 gives a factor of about 120 enhancement (if one
keeps
the same other quantum numbers)!

 From these expressions we see that present experimental limits
\cite{L3,Delphi},
$Br(Z\ra 3\gamma)\sim 10^{-5}$ place a bound on
the strength of the $Z3\gamma$ of the order of
\beq
4 \left( \frac{\alpha M_Z^2}{M^2} \right)^2 \tilde{\beta}_{2,3} \sim 0.4.
\eeq
Therefore, one can easily argue that a  conventional weakly interacting
particle
($\tilde{\beta} \sim 1$)
can not be probed through the $Z\ra 3\gamma$ decay. This is not to say that we
should not look for these decays; this would be like arguing that lacking any
model that gives large $WW\gamma/Z$ anomalous coupling one should not probe
the $WW\gamma/Z$ couplings at LEP2!

The single-photon energy distribution is
\beqn
\frac{1}{\Gamma} \frac{{\rm d}\Gamma}{{\rm d}x}=\frac{20}{3\bzpc + 5 \bzmc}
x^3 \left( (\bzpc +  \bzmc) (1-x) + \frac{\bzpc + 2\bzmc}{5} (20-40x+21 x^2)
\right).
\eeqn
It is more telling and appropriate to distinguish the photons by
ordering them according to their energy; we will label  them as
$E_1>E_2>E_3$ ($E_i$ is the energy).  The energy distribution of the least
energetic photon is
\beqn
\label{dise3analytical}
\frac{1}{\Gamma} \frac{{\rm d}\Gamma}{{\rm d}x_3} 
&=&\frac{12}{3\bzpc + 5 \bzmc} \nonumber \\
& &\left\{ x_3^3 \left[(\bzpc + 2 \bzmc) (1+3 x_3 (1-x_3)+ 24 (1-x_3)^2)-5\bzmc
(1-x_3) \right]
\theta(\frac{1}{2}-x_3) \right. \nonumber \\
&+&  (2-3x_3)  \left[
(\bzpc + 2 \bzmc) (1+3 (1-x_3)(1+(1-x_3)(4-23x_3+27x_3^2)) \right. \nonumber \\
 &-&    \left. \left. 5\bzmc (1-x_3)(1-3 (1-x_3)^2) \right]
\theta(2/3-x_3)\theta(x_3-1/2)
\right\},
\eeqn
where $\theta(x)$ is the step function.

The average energy of the least energetic photon, $E_{3}$, is
\beqn
<E_{3}>=\frac{M_Z}{42} \frac{1}{3 \bzpc+5\bzmc} \left(\frac{751}{27}
\bzpc+\frac{367}{8} \bzmc \right).
\eeqn
This shows that independently of $\bzm,\bzp$ the average energy of
the photon is almost the same for the two independent operators, {\it i.e.},
$0.221 M_Z$ for $\bzp$ and $0.218$ for $\bzm$. As expected these are hard
photons as compared to the  expected average energy  of the softest photon one
gets for a typical
QED $3\gamma$ event. Therefore,  it is relatively easy to disentangle the
QED and  \z3gt events; however it is not easy to distinguish between
different types of New Physics. This
will be made clearer when we show the results of the full simulations including
cuts. It is also difficult to differentiate between the models on the basis of
the average energy of the most energetic photon and the ``medium" one.

It is relatively straightforward to implement the cuts within the event plane
as done in the LEP experiments. Requiring
the separation angle between any two photons ($\theta_{ij}$) to be above a
certain value and requiring a cut on the energy of the photons, such that
\beqn
\label{theocuts}
\theta_{ij}> 2 \sqrt{\mbox{\rm Arcsin} \;\delta}\;\;\; ; \;\;\;
x_i=\frac{2E_i}{M_Z}> \varepsilon,
\eeqn
we calculate the acceptance to be

\pagebreak

\beqn
\label{approxcutsg}
\frac{\Gamma_{{\rm cut}}}{\Gamma}-1\simeq -\frac{1}{3\bzpc+5\bzmc}\left(
(\bzpc+2\bzmc)\;\frac{18 \delta}{7}+
(\bzpc+\bzmc) \; \frac{9\delta^2}{7} + (4\bzpc+5\bzmc)\;\frac{2 \delta^3}{7}
\right.
\nonumber \\
\left. \hspace*{3cm}  \frac{}{}+
(5\bzpc+9\bzmc)\;  15\varepsilon^4 \right). \nonumber \\
\;\;\;
\eeqn

We refrain from giving the exact expression, not only because it is lengthy
and uninspiring but also because the typical values of the cuts applied by
the LEP experiments are such that  the above approximate expression is
more than enough.
Note for
instance that the cut on the energy is of order $\varepsilon^4$ reflecting once
again the preferentially energetic photon characteristics of the New Physics
and
the quartic dependence of the differential width on the energy. This is the
reason we have  kept terms up to  fourth order in the
cut-offs.  At this order there is ``no-mixing"
between the cuts on the opening angles and  the energy.
With $E_i>M_Z/20$ and $\theta_{ij}> 20^{0}$ (See
OPAL  \cite{Opal}),
one does not lose more than a few percent of the signal.
The message then is that one can afford applying harder cuts than the above
without much affecting the signal; even with cuts that are twice as large as
the
typical LEP experiments cuts our approximate formula, Eq.~\ref{approxcutsg}, is
excellent. The main purpose of the ``mild" LEP1
experimental cuts is to bring about a
decent and clean $3\gamma$ sample (which in the absence of any new physics is
purely QED). With higher statistics we could afford cutting
harder. It is worth emphasising that the acceptance is sensibly the same for
different manifestations of New Physics, {\it i.e.}, whether one takes
\bzpt$\;$ or
\bzmt$\;$. Again it has to do with the fact that the dependence on the photon
energies for different choices of $\tilde{\beta}$ are not terribly different.

Introducing the correlations with the initial state, we have been able to
obtain some
exact results including cuts. For the sake of generality we also give
the cross section for polarised electron beams.

For the $3 \gamma$ final state we have that
\beqn
\sigma(e^+e_{L,R}^{-}\stackrel{Z}{\ra} 3\gamma)=2 \;
\frac{g_{L,R}^2}{g_L^2+g_R^2}
 \sigma_{{\rm unp.}}
\eeqn
where $\sigma_{{\rm unp.}}$ is the unpolarised cross section
and
\beq
g_L=-\frac{1}{2}+s_W^2\;\;\;\;\;\;\;;\;\;\;\;\; g_R=s_W^2.
\eeq
We introduce, for a centre-of-mass energy, $\sqrt{s}$, the $Z$ propagator via
\beq
D_Z(s)=\frac{(s-M_Z^2)-i\Gamma_Z M_Z}{(s-M_Z^2)^2+(\Gamma_Z M_Z)^2}
\eeq
where we have chosen to keep an energy-independent width. In terms of the
physical width of the $Z$ into electrons, $\Gamma_Z^{ee}$, and with
$dLips$ being the invariant 3$\gamma$ phase space,
the differential cross section writes as
\beqn
&&\frac{1}{90\pi}\frac{{\rm d}\sigma(\epem \stackrel{Z}{\ra}
3\gamma)}{(16\pi)^2{\rm d}Lips}=
 12 \pi \frac{ \Gamma_Z^{ee} \Gamma_Z^{3\gamma}}{M_Z^2}
\left( \frac{s}{M_Z^2} \right)^4 |D_Z(s)|^2
\Bigg\{  \nonumber \\
& & \frac{\bzpc+\bzmc}{3\bzpc+5\bzmc} \left[
2 (1-x_1)(1-x_2)(1-x_3) -\left( x_1^2 c_1^2 \;(1-x_1-\frac{x_2 x_3}{2}) \right.
\right. \nonumber \\
& &\;\;\;\;\;\;\;\; \;\;\; \left. \left.
\;\;\;\;\;+\;\;\;\;x_2^2 c_2^2 \;(1-x_2-\frac{x_1 x_3}{2}) + x_3^2 c_3^2
\;(1-x_3-\frac{x_1 x_2}{2})
\right) \right]  \nonumber \\
&+&\frac{\bzpc+2\bzmc}{2 (3\bzpc+5\bzmc)}  \left[
\left( (1-x_1)^2 (1-x_2)^2 + (1-x_1)^2 (1-x_3)^2+ (1-x_2)^2 (1-x_3)^2
\right. \right.  \nonumber \\
&& \;\;\;\;\;\;\;\;\; \;\;\;\;\;\;\;\;\;\;\;\;\;\;\;
\left. -\;3 (1-x_1)(1-x_2)(1-x_3) \right)   \nonumber \\
&+& \left.\frac{1}{2}
\left(
x_1^2 c_1^2 \; (1-2 x_1+x_2^2 +x_3^2) +
x_2^2 c_2^2 \; (1-2 x_2+x_1^2 +x_3^2) +
x_3^2 c_3^2 \;(1-2 x_3+x_1^2 +x_2^2)  \right) \right] \Bigg\}. \nonumber \\
& &\;\;\;
\eeqn
Here $c_i$ is the cosine of the angle between photon $i$ and the beam,
$c_i=\cos (e^-\gamma_i)$ and $x_i=E_i/\sqrt{s}$.

 From the above expression we can look at the orientation of the photons
relative to the beam. For instance,  the distribution in the
angle $\psi$ between the event
plane and the beam axis, with $z=\sin(\psi)$ ($0<\psi<\pi/2$)  , is given by
\beqn
\frac{1}{\sigma} \frac{{\rm d}\sigma}{{\rm d}z}=\frac{27}{8}
\frac{\bzpc+2\bzmc}{3\bzpc+5\bzmc}\left\{
1-\frac{z^2}{9}\; \frac{3\bzpc+14\bzmc}{\bzpc+2\bzmc} \right\}.
\eeqn
When specialising to the ``fermion-monopole" case, we disagree with
  \cite{DeRujula3g}. In principle it is this distribution that can
disentangle between different $\tilde{\beta}$.
The average value of $z$ may be easily computed
\beq
<z>=\frac{3}{32} \frac{15\bzpc+22\bzmc}{3\bzpc+5\bzmc}.
\eeq
The totally integrated cross section at a given energy writes very simply in
terms of the
width measured at the $Z$ resonance:
\beqn
\sigma^{NP}(\epem \ra 3\gamma)= \frac{12 \pi}{M_Z^2}\;
 \Gamma_{Z}^{\epem}\; \Gamma_{Z}^{3\gamma} \;\frac{1}{s}
 \left( \frac{s}{M_Z^2} \right)^4 |s D_Z(s)|^2.
\eeqn
This shows that, as expected, the cross section  grows like $s^3$.
Hence, at some energy much above the $Z$ peak this growth factor can  give
an enhancement factor of the order of that given by the resonance (see below).

Of course at the $Z$ peak one has the usual Breit-Wigner formula:
\beqn
\sigma^{NP}(\epem \ra 3\gamma)= \frac{12 \pi}{M_Z^2}
 \frac{\Gamma_{Z\ra \epem} \Gamma_{Z \ra 3\gamma}}{\Gamma_Z^{{\rm tot}}}.
\eeqn

It is also possible to give an exact expression for
the $s$-channel cross section including realistic cuts. Besides the
cuts on the photon energies and photon-photon separations (see
Eq.~\ref{theocuts}), we can include
a cut on the angle, $\psi$,  between the event plane and the beam to avoid
 forward-backward events and reduce the purely QED contribution such that
$\sin(\psi)>\sin(\psi_0)=z_0$.
The exact analytical formula is lengthy and not very telling. With
the typical  values of the experimental cuts as applied by the LEP
collaborations
the exact formula is well approximated by:
\beqn
\frac{\sigma_{{\rm cuts}}}{\sigma}-1\simeq
\frac{1}{3\bzpc+5\bzmc} \left\{15 (5\bzpc+9\bzmc)\varepsilon^4
-18 (\bzpc+2\bzmc) (\frac{1}{7} \delta +
\frac{3}{16} z_0) \right\} \nonumber \\
\;\;\;\;
\eeqn

\setcounter{subsection}{0}
\setcounter{equation}{0}
\def\thesubsection {\thesection.\arabic{subsection}}
\def\theequation{\thesection.\arabic{equation}}
\def\thefigure{\arabic{figure}}
\section{Monte Carlo Analysis}

\subsection{Behaviour of the cross section and effect of the interference}

The above expression can prove handy for a quick estimate of the
efficiencies on the detection and for the measurement of the
$Z\ra 3\gamma$ partial width. However, to conduct a detailed analysis
it is essential to consider the QED $3\gamma$ background as well as the effect
of
the interference and be able to apply any cut that is needed for a particular
detector.
To include more sophisticated
cuts a MC program is necessary.
We have written a fast MC which includes both the pure
QED contribution with the 3 {\em observable} photons as well
as the anomalous terms. The program is based on a calculation of the
matrix elements. This allows us to study the effect of the interference
between the standard part and the non-QED part. The very
compact expressions are derived through the use of the spinor inner
product  \cite{GastmansWu,Manganophysrep,Mangano,Kleiss,XZC} and can be found
in the
Appendix. By implementing the same
cuts that we implemented in the analytical formulae we find
excellent agreement between the analytical results and the output
of the MC for all the parts  contributing to the cross section.
We will explicitly show that the interference term can always
be dropped, hence simplifying the experimental analysis even when one is far
away
from the peak.

Of course at the $Z$ peak, there is no interference since the tree-level
QED amplitude is purely real whereas the anomalous is purely imaginary. We find
that if the search  at the $Z$ peak reveals that the anomalous part is smaller
than the QED part (which is the case) this implies values for the couplings
such that even at LEP2 the interference is negligible.
Moreover, it is only the term in $\tilde{\beta}_+$ that contributes to
the interference, the reason being that the tree-level
cross section for the production of 3 photons with the same helicity is known
to be exactly
zero \cite{GastmansWu}\footnote{This also holds for
same-helicity multi-photon final state processes  \cite{Mangano}.}. Since, as
remarked in
the first  section, the classification into $\bzp$ and $\bzm$ refers to two
classes of helicities for the photons, only $\bzp$ interferes. Details are
found in the Appendix.

It is always possible to give a parameterisation for the cross section as a
function of the cms energy, including any cuts. A very simple scaling law
parametrisation is obtained if the angular cuts are the same at all energies
and
if the  cut on the photon energies are implemented through a fixed ratio
with the total cms energy. In an obvious notation
\begin{figure*}[htbp]
\begin{center}
\caption{\label{interference}{\em Relative effect of the interference for
various values of
$\tilde{\beta}_+$. Positive $\tilde{\beta}_+$ were taken. We have preferred,
instead of quoting the values of the $\tilde{\beta}_+$, to give the associated
 $3\gamma$ width  as calculated from Eq.3.2.}}
\vspace*{0.5cm}
\mbox{\epsfxsize=16cm\epsfysize=9cm\epsffile{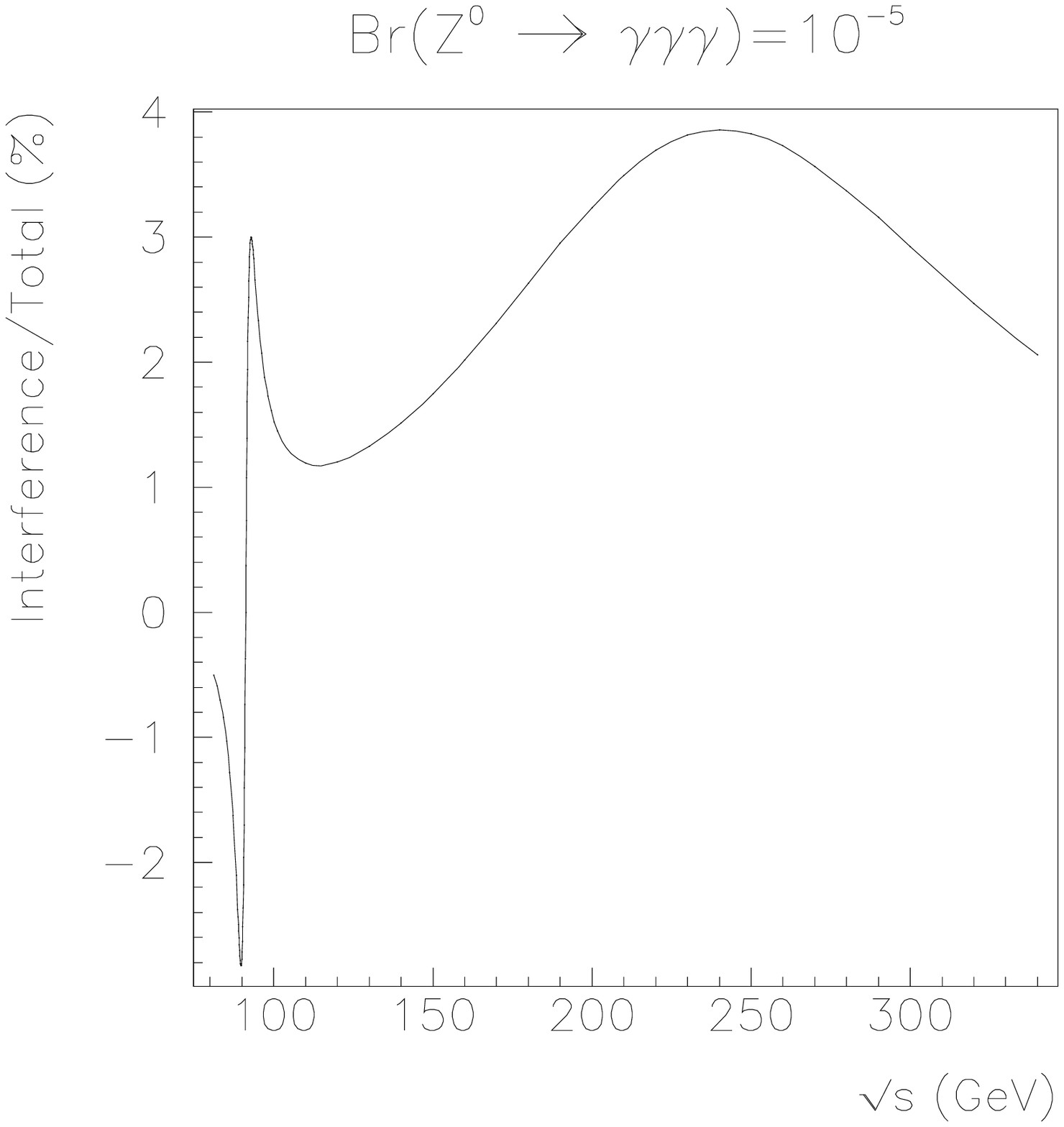}}
\vspace*{0.5cm}
\mbox{
\mbox{\epsfxsize=7.5cm\epsfysize=9cm\epsffile{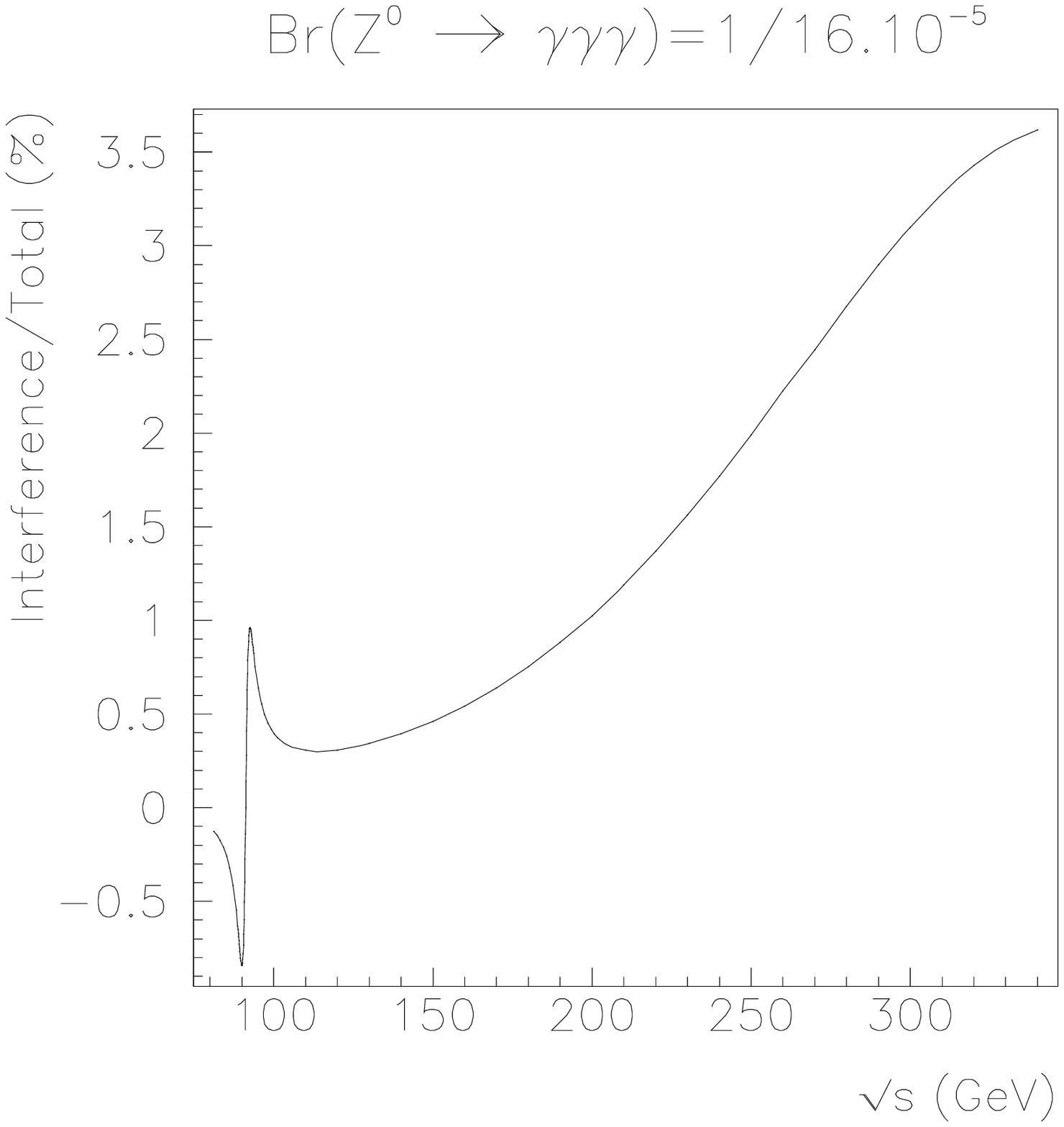}}
\hspace*{0.5cm}
\mbox{\epsfxsize=7.5cm\epsfysize=9cm\epsffile{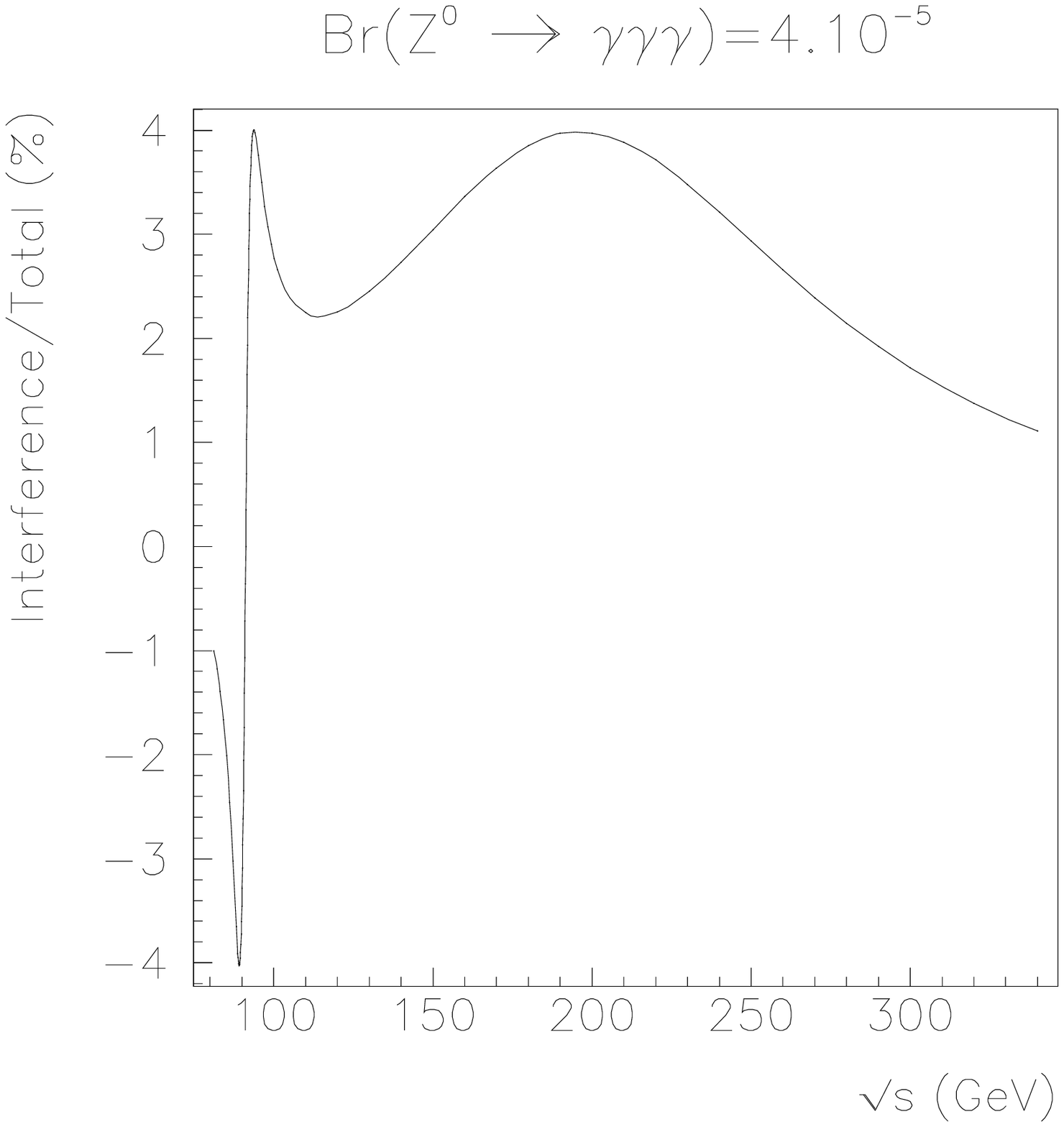}}}
\vspace*{-2cm}
\end{center}
\end{figure*}
\beqn
\label{ee3gsigs}
\sigma(\epem \ra 3\gamma)&=&\sigma_0 \frac{1}{\tilde{s}} \;\;
+\frac{4\alpha^2}{m^4} \; \tilde{\beta}_+ \;  \sigma_i \;
\frac{(\tilde{s}-1)\tilde{s}^2 }{\displaystyle
\left(\frac{\tilde{s}-1}{\gamma}\right)^2+1}
\nonumber \\
&+&(\frac{4\alpha^2}{m^4})^2 \;(\bzpc \sigma_+ + \bzmc \sigma_-)
 \frac{\tilde{s}^5}{\displaystyle \left(\frac{\tilde{s}-1}{\gamma}\right)^2+1},
\eeqn
\beqn
\hbox{with} \;\;\;\;\;\; \tilde{s}&= &\frac{s}{M_Z^2} \;\;\;\; ; \;\;\;\;\;\;
\gamma=\frac{\Gamma_Z}{M_Z} \;\;\;\; ; \;\;\;\;\;\ m=\frac{M}{M_Z}.
\eeqn
For instance, taking the OPAL cuts  \cite{Opal}
\beqn
\label{expcuts}
|\cos\theta_{e\gamma}|<0.9 \;\;\;\;; \;\;\;\; \theta_{\gamma \gamma}>20^{0}
\;\;\;\;; \;\;\;\;
E_\gamma> \frac{\sqrt{s}}{20}
\eeqn
the values of the cross section $\sigma_{0,i,+,-}$ calculated with
$M_Z=91.187$~GeV, $\Gamma_Z=2.490$~GeV, $\sin^2\theta_W=s_W^2=0.232$, are
\beqn
\label{sigvalues}
\sigma_0&=&  0.705\; \mbox{\rm pb}    \nonumber \\
\sigma_i&=& 0.082 \; \mbox{\rm pb}     \nonumber \\
\sigma_+&=&  4.181 \; \mbox{\rm pb}   \nonumber \\
\sigma_-&=&   6.420 \; \mbox{\rm pb}.
\eeqn

Armed with the MC, simulating an experimental environment, we can first exactly
quantify the effect of the interference. We only consider models with
$\tilde{\beta}_+$
since
there is no interference with $\tilde{\beta}_-$.
As a first illustration we take a value of $\tilde{\beta}_+/M^4$ which
corresponds to the best limit set on the branching ratio for \z3gt \cite{L3}:
$Br(Z\ra 3\gamma)\sim 10^{-5}$. We clearly see (Fig.~\ref{interference}) that
around the
energies scanned by LEP1, the effect is negligible. The relative error
introduced by neglecting the interference  is below 3$\%$  all the
way up to LEP2 energies. Even for an energy around 300~GeV the effect does not
exceed $3\%$.
We have also looked at the effect of the interference for a value of
$\tilde{\beta}_+$ which is 4 times
smaller (that would give a branching fraction 16 times smaller than the best
published limit and which could be considered as the ultimate limit from LEP1).
The effect up to LEP2 energies is below the 1$\%$ level. We have also
considered values that are  20 times smaller; however, foreseeing the
accumulated
luminosity at LEP1 it would not be possible to detect the effect of an
anomalous
\z3gt at LEP. Nonetheless, even for such values the effect of the interference
is
too small even at LEP2 (and beyond) energies. Just to make the point we also
show
that there is negligible effect even when taking  larger values of the
coupling (Figs~\ref{interference}).
\begin{figure*}[htbp]
\begin{center}
\caption{\label{curve}{\em Energy dependence of the cross section for various
values of
$\tilde{\beta}_+$. The full thick curve is the total contribution including the
interference while the dotted curve is with no interference. Also shown,
separately, the irreducible QED cross section.}}
\vspace*{0.5cm}
\mbox{\epsfxsize=16cm\epsfysize=9cm\epsffile{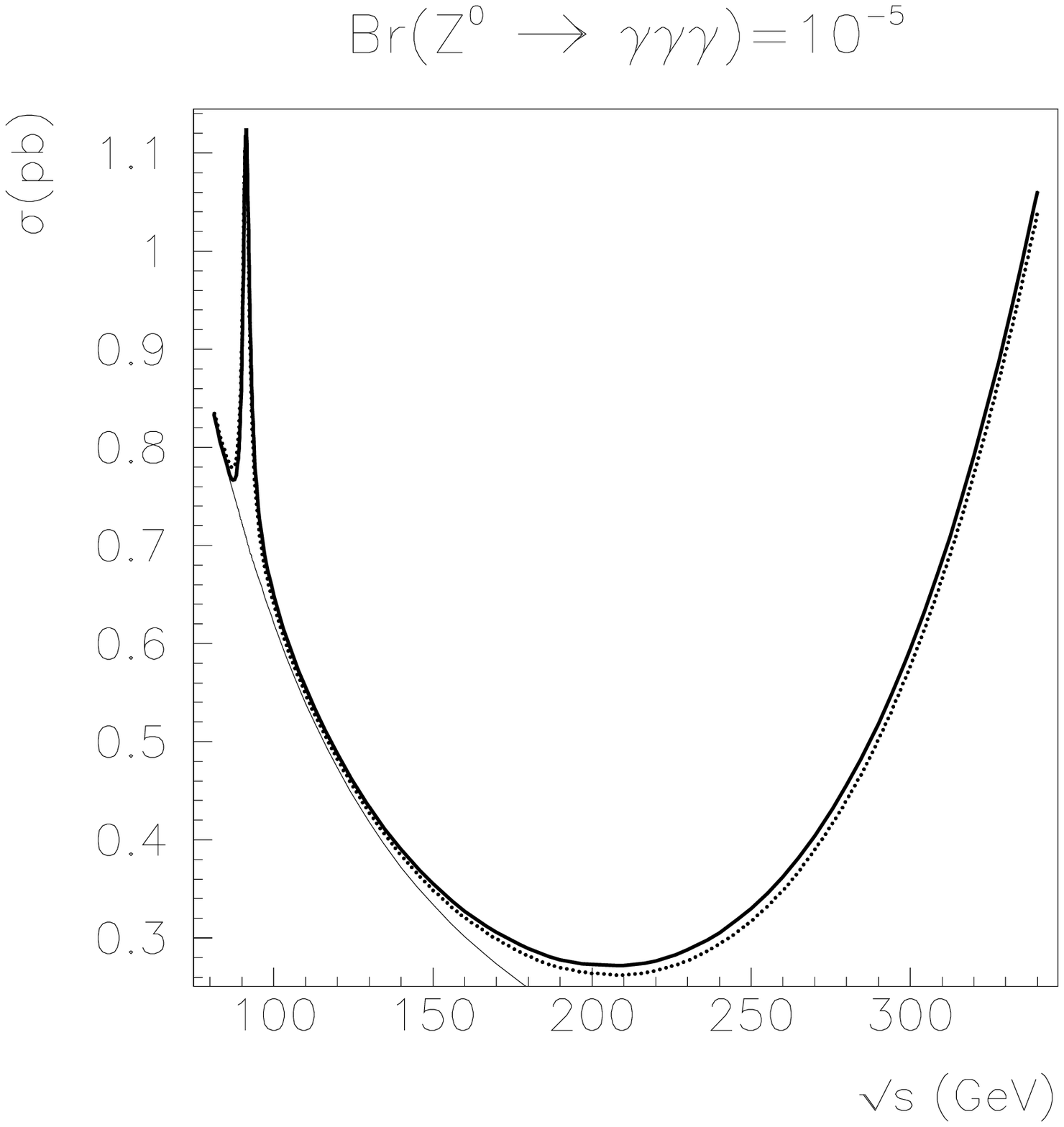}}
\mbox{
 \mbox{\epsfxsize=7.5cm\epsfysize=9cm\epsffile{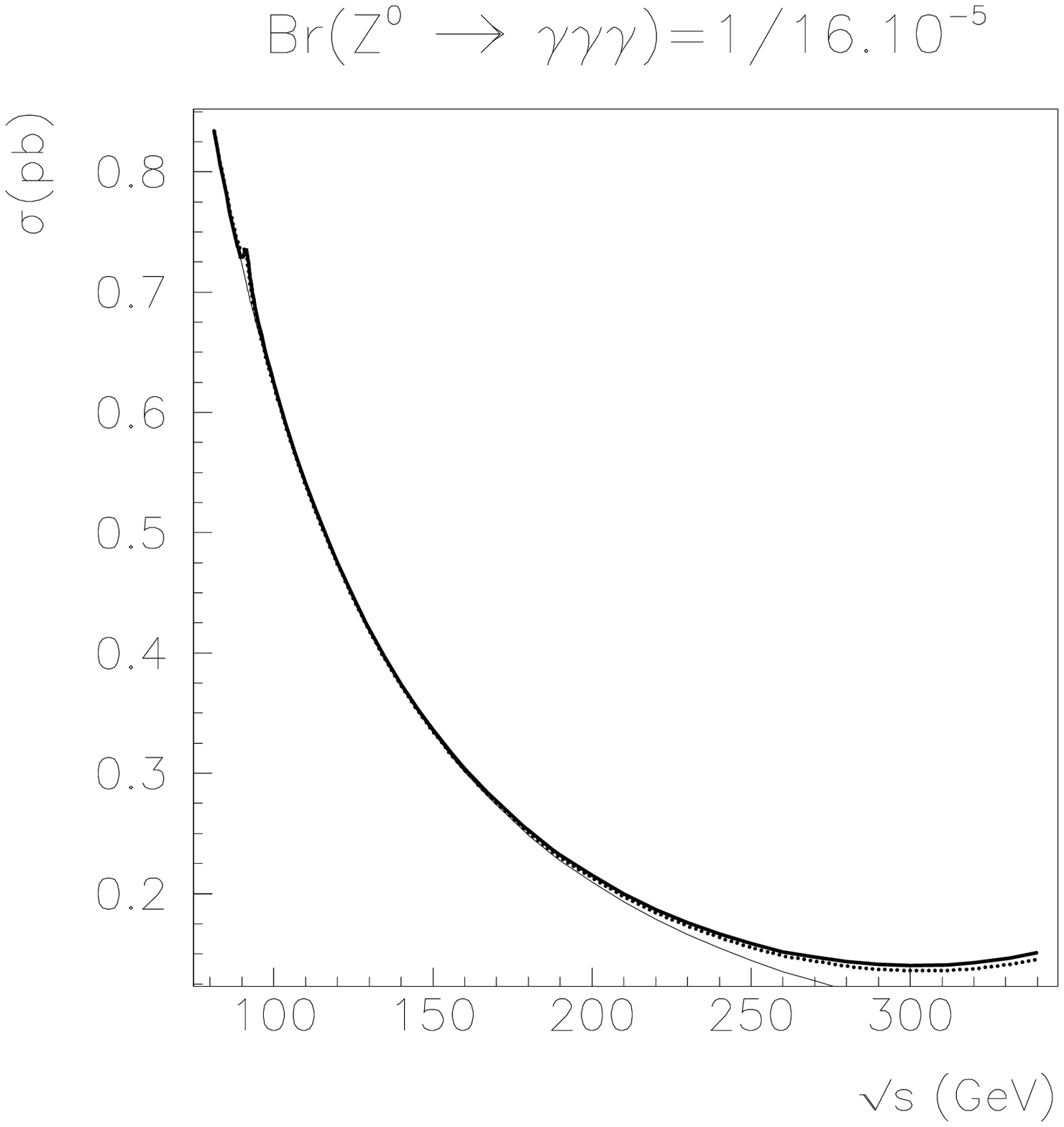}}
\hspace*{.5cm}
\mbox{\epsfxsize=7.5cm\epsfysize=9cm\epsffile{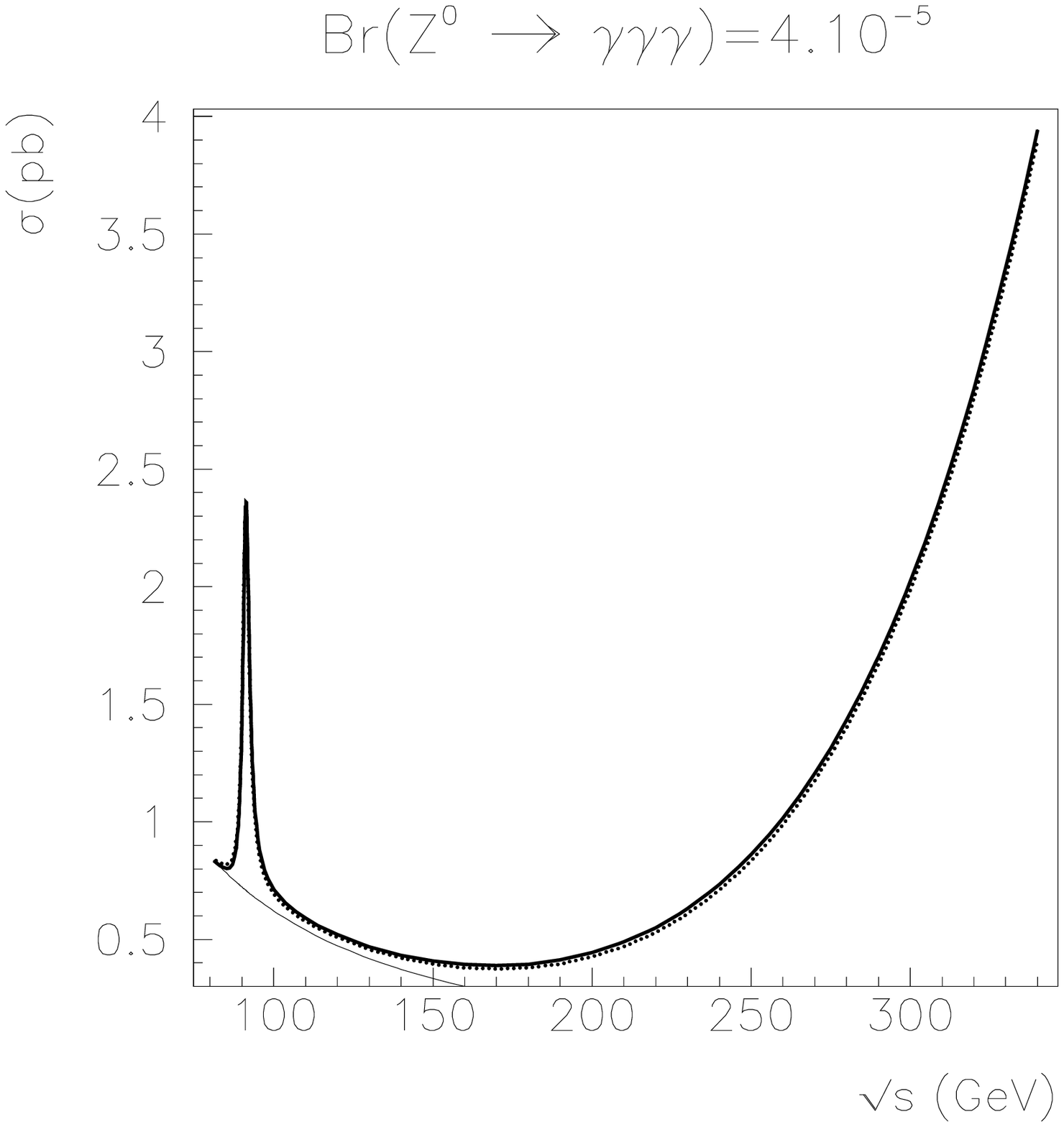}}}
 \vspace*{-2cm}
\end{center}
\end{figure*}

Figures~\ref{curve} show the merit of sitting at the peak contrasted
to the improvement one gains by moving to higher (than LEP2) energies, taking
for $\tilde{\beta}$ typical values that one may hope to measure at LEP1.
We see that just above the $Z$ peak the
combined cross section
(QED+anomalous) decreases (due to the Breit-Wigner factor and the 1/$s$ drop of
the QED cross section) and that it starts bending over and increasing only,
unfortunately, around the highest LEP2 energies. One
conclusion, to which we will come back later, is that these figures convey the
message that if there is no sign of New Physics affecting the \z3gt at LEP1
even the highest energies foreseen for LEP2 will not be enough to probe this
vertex better, even with the higher nominal luminosity of LEP2 (500~$pb^{-1}$).

\subsection{Distributions}
The selection of the optimal cuts that should be applied in order to bring out
a
\z3gt effect should be
guided by the knowledge of the various distributions in energies and angles.
These are best given by the MC. We have looked at a few distributions
applying on both
the QED and anomalous parts the same cuts as defined in the previous
subsection,
see Eq.~\ref{expcuts}.
We have chosen to
show separately the {\em normalised} (to one) cross section for the anomalous
and the QED; the effect of the interference is not taken into account.
However we have seen that the interference can be safely neglected; moreover,
in
terms of the parameter $\bzm$, this interference is not present, as it is not
present at
all on the $Z$ peak for all operators.  Beside the variables we discussed
earlier we also
looked at the
distribution in the acollinearity in the transverse plane
 defined by the most energetic photon (1) and
the ``medium" photon (2). This is defined from the photon momenta transverse to
the
beam, $\vec{k}_{1,2}^T$ as
\beq
\chi= \mbox{\rm Arcos}\left(\frac{ \vec{k}_{1}^T .
\vec{k}_{2}^T }{|\vec{k}_{1}^T|\;|\vec{k}_{2}^T|}  \right). \nonumber
\eeq
For all the distributions we show the effect of $\bzm$ and $\bzp$ and one of
the
other ``models"
defined in the second section in order to see how these models may be
disentangled, in
principle. For the single distribution variables we also show the average value
of the corresponding variable including the cuts for the QED as well as the
anomalous.

We start with the energy distributions. As argued above there is a
clear-cut distinction between the QED and the anomalous contribution that is
best understood for the least energetic photon.
\begin{figure*}[htbp]
\begin{center}
\caption{\label{dise3}{\em
Normalised distribution in the energy of the softest photon for the
case of QED as well as for the models defined in the text. Cuts as defined by
Eq.~\protect\ref{expcuts} have been applied. The arrows point to the average
values of the energy of the softest photon for the models considered. }}
\vspace*{ 1.0cm}
 \mbox{\epsfxsize=16.cm\epsfysize=16cm\epsffile{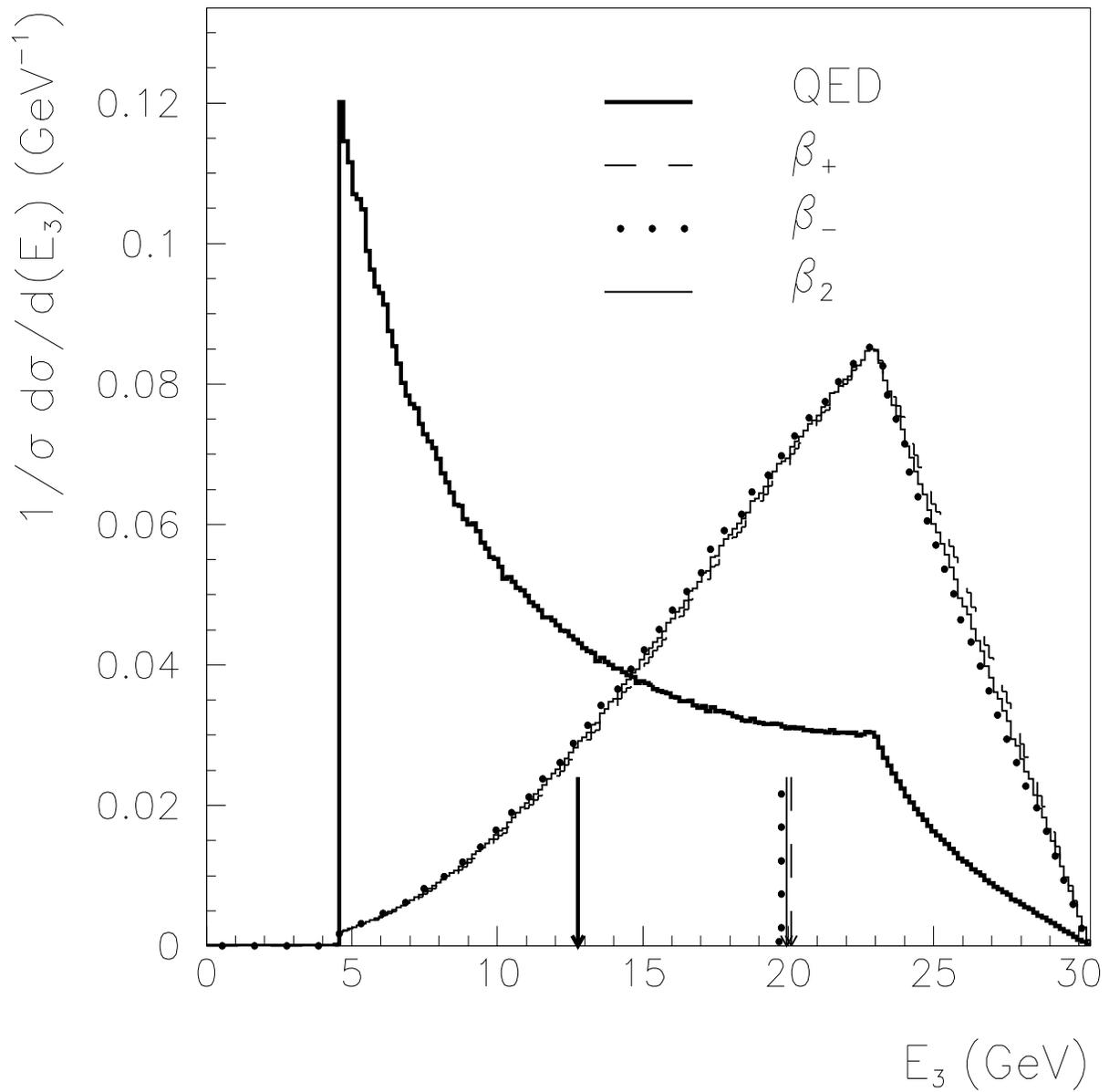}}
\end{center}
\end{figure*}
 In QED $e^+ e^- \ra 3\gamma$ is built up
from
$e^+ e^- \ra 2\gamma$ where the ``added" photon can be considered to be a
bremstrahl. One should, therefore, expect that among the three photons, one is
much less
energetic than the other two (that  will tend to take the beam energy). In the
case of the
$Z\rightarrow 3\gamma$, all three photons have on average an energy of the same
order, thus the energy is shared somehow equally between them. This is the
reason one of
the most salient differences appears in the distribution of the least
energetic $\gamma$. This is well rendered by Fig.~\ref{dise3} where two very
distinctive
spectra stand out. This distribution is thus a powerful tool for separating the
QED from the anomalous. Note, however, that this distribution is almost
insensitive to
the model of New Physics contributing to the $s$-channel \z3gt. Therefore based
on
this remark one could devise a ``model-independent" optimised cut to unravel
the
anomalous contribution\footnote{The ``knee" that shows up at $M_Z/4$ is purely
kinematical and has to do with the ordering of the photons. This effect is also
manifest in the analytical formula Eq.~\ref{dise3analytical} through the
occurrence of the step
function at $x_3=1/2$.}.
A similar discrimination between the anomalous and the QED occurs in the case
of the
medium-energy photon without ``resolving" the models.  The distribution in the
hardest photon shows a less dramatic difference (see Figs.~\ref{dise12}).
\begin{figure*}[htbp]
\begin{center}
\caption{\label{dise12}{\em As in the previous figure for the medium photon,
$E_2$ and the hardest, $E_1$.}}
 \vspace*{0.5cm}
 \vfill
 \mbox{\epsfxsize=10.5cm\epsfysize=9.cm\epsffile{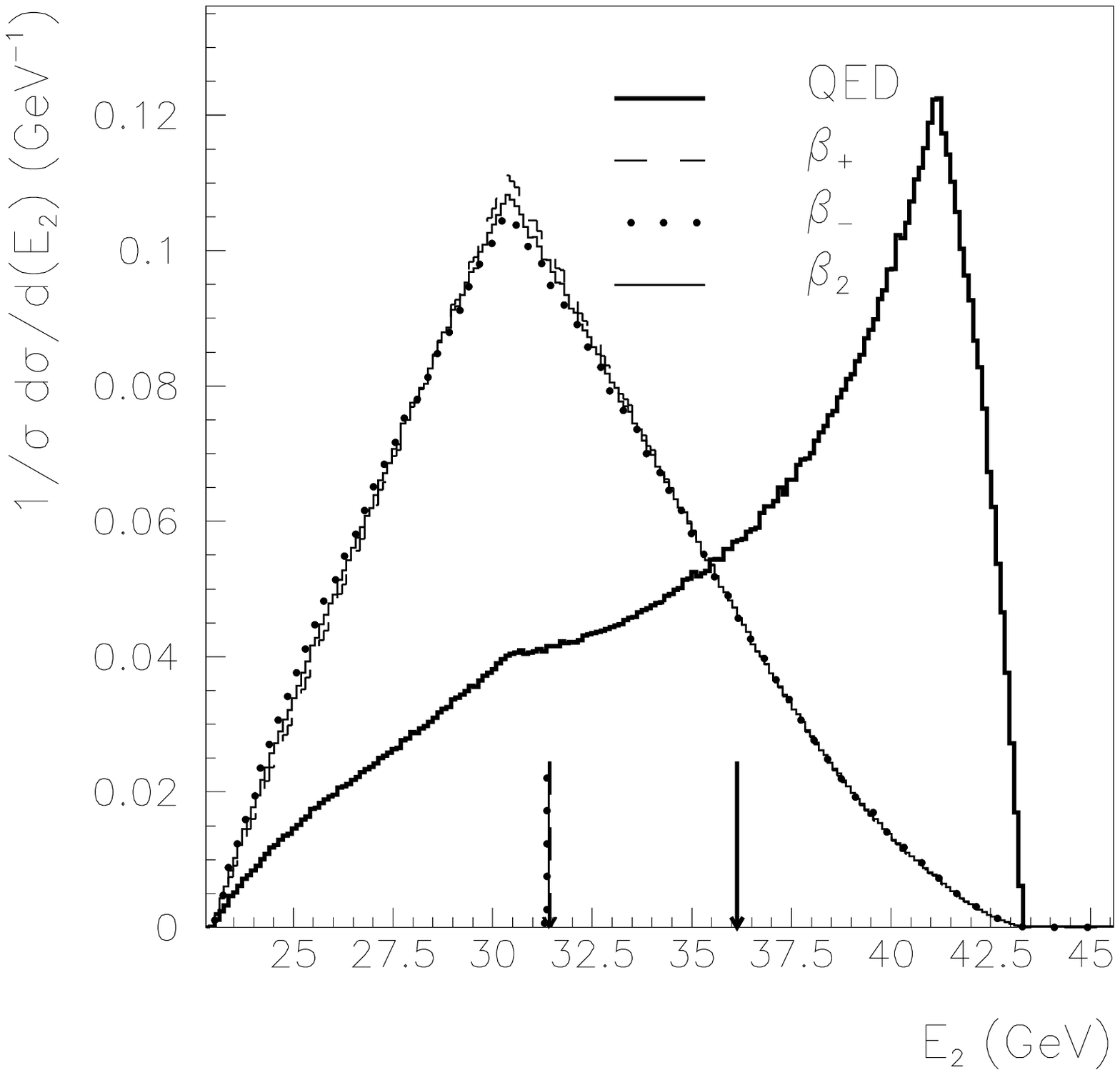}}
 \vspace*{1.0cm}
 \vfill
 \mbox{\epsfxsize=10.5cm\epsfysize=9.cm\epsffile{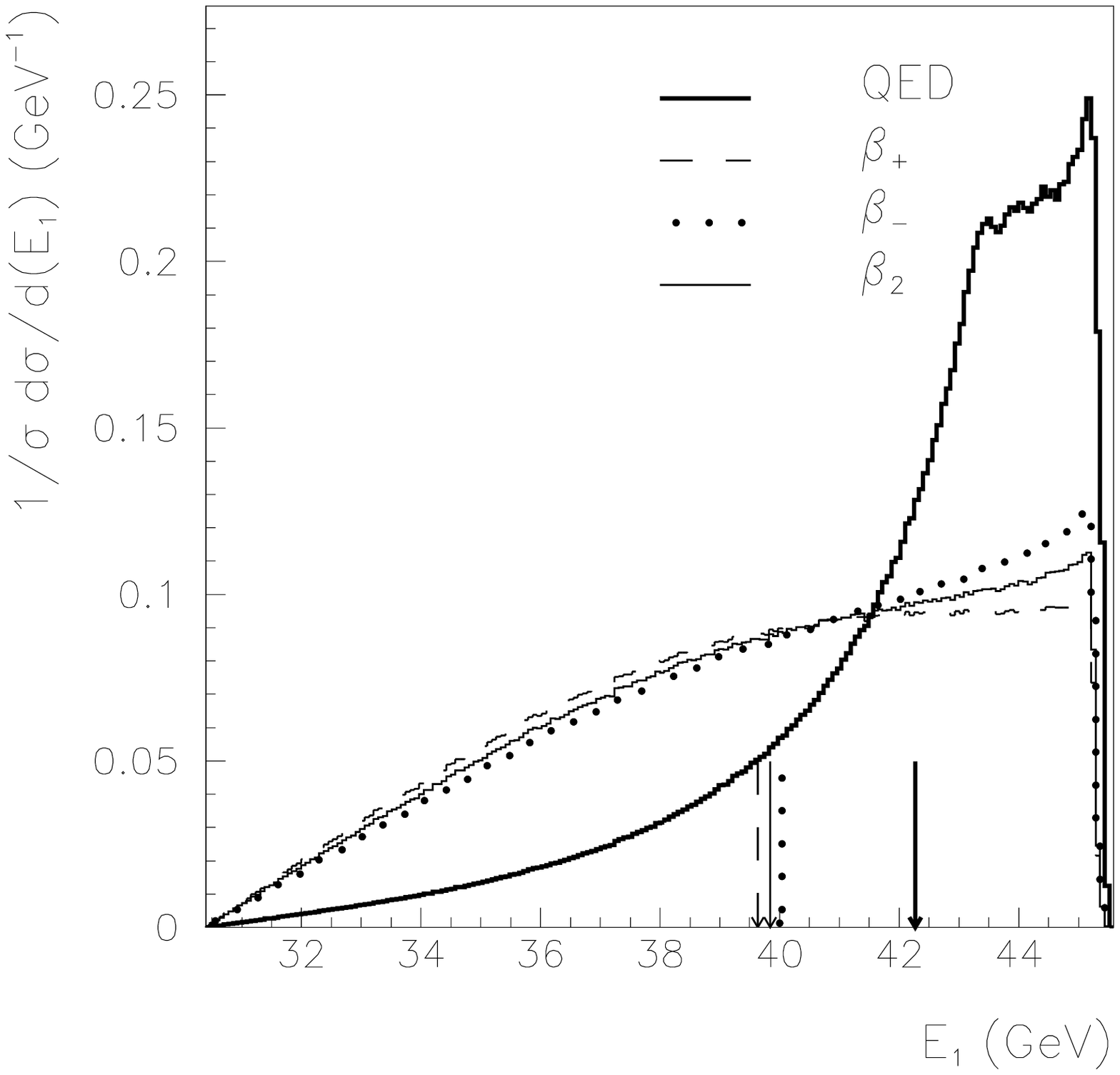}}
 \vfill
\end{center}
\end{figure*}
Other distributions in the event plane variables relate to the opening angles
between any two photons.
\begin{figure*}[htbp]
\begin{center}
\caption{\label{disoij}{\em As in Fig.~\protect\ref{dise3} but in the opening
angles between
the photons. ``3" is the softest and ``1" the hardest.}}
 \vspace*{0.5cm}
 \mbox{\epsfxsize=14cm\epsfysize=6.3cm\epsffile{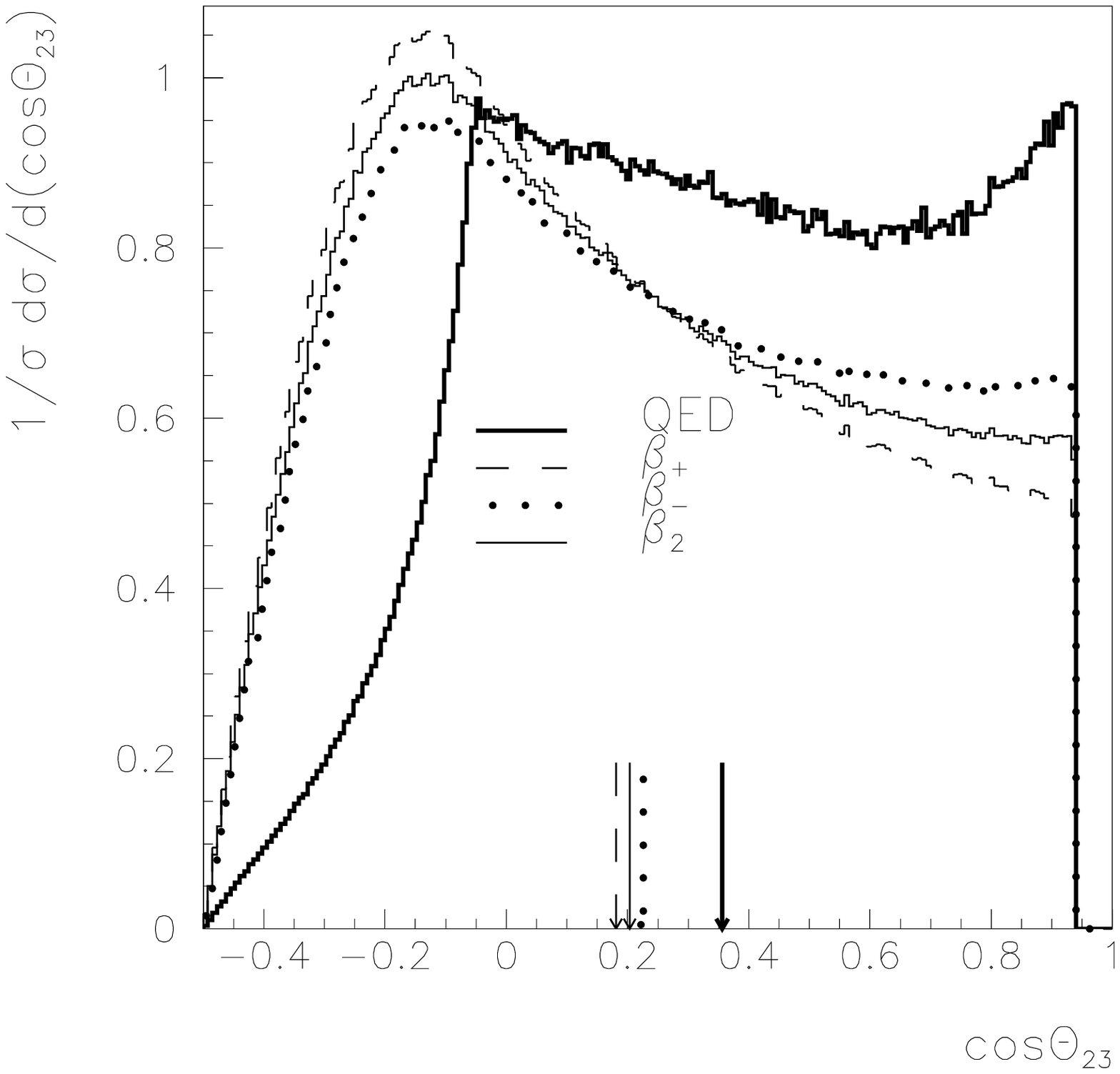}}
 \vspace*{0.5cm}
 \mbox{\epsfxsize=14cm\epsfysize=6.3cm\epsffile{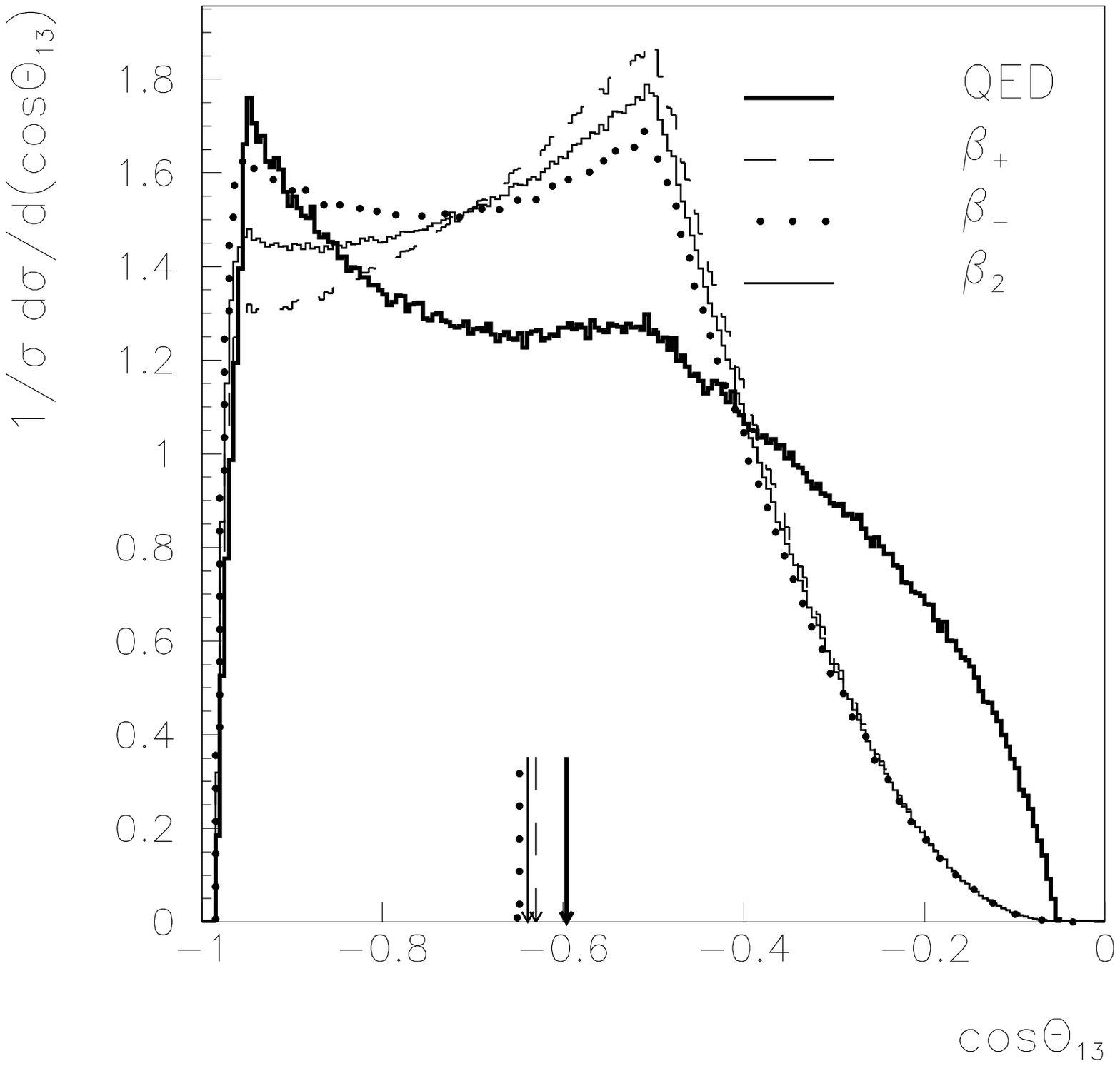}}
 \vspace*{0.5cm}
 \mbox{\epsfxsize=14cm\epsfysize=6.3cm\epsffile{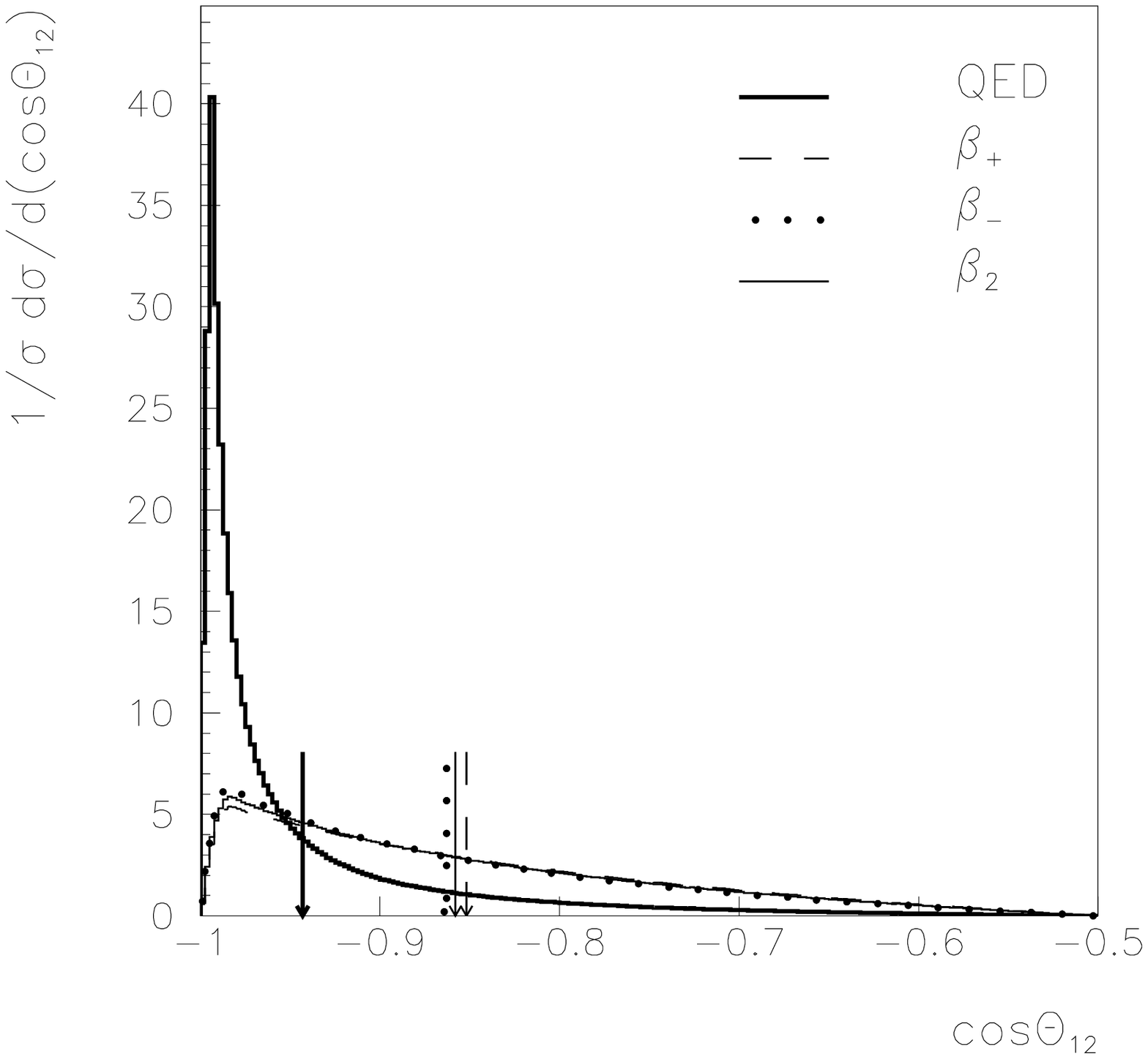}}
\end{center}
\end{figure*}
Although these distributions are directly
related to the energies ($\sin^2(\theta_{ij}/2)=(x_i+x_j-1)/x_i x_j$),
besides bringing out the markedly different structure of the QED and
the anomalous, one finds that given enough non-QED events a certain
discrimination may be perceptible, especially in the distribution of the
angle between the two softest,
Figs.~\ref{disoij}.

Another characteristic of QED events, besides the softness of one of the
photons,
is the collinearity
of the photon with the particle that emits that photon. This is not expected
for
the \z3gt where the photon does not connect to the \epm line. Thus another
clear signature is the distribution in the angle between any photon and the
beam.
\begin{figure*}[htbp]
\begin{center}
\caption{\label{disfbeam}{\em As in Fig.~\protect\ref{dise3} but in the cosine
of the angle between the photons and the beam.}}
 \vspace*{.5cm}
 \mbox{\epsfxsize=14cm\epsfysize=6.3cm\epsffile{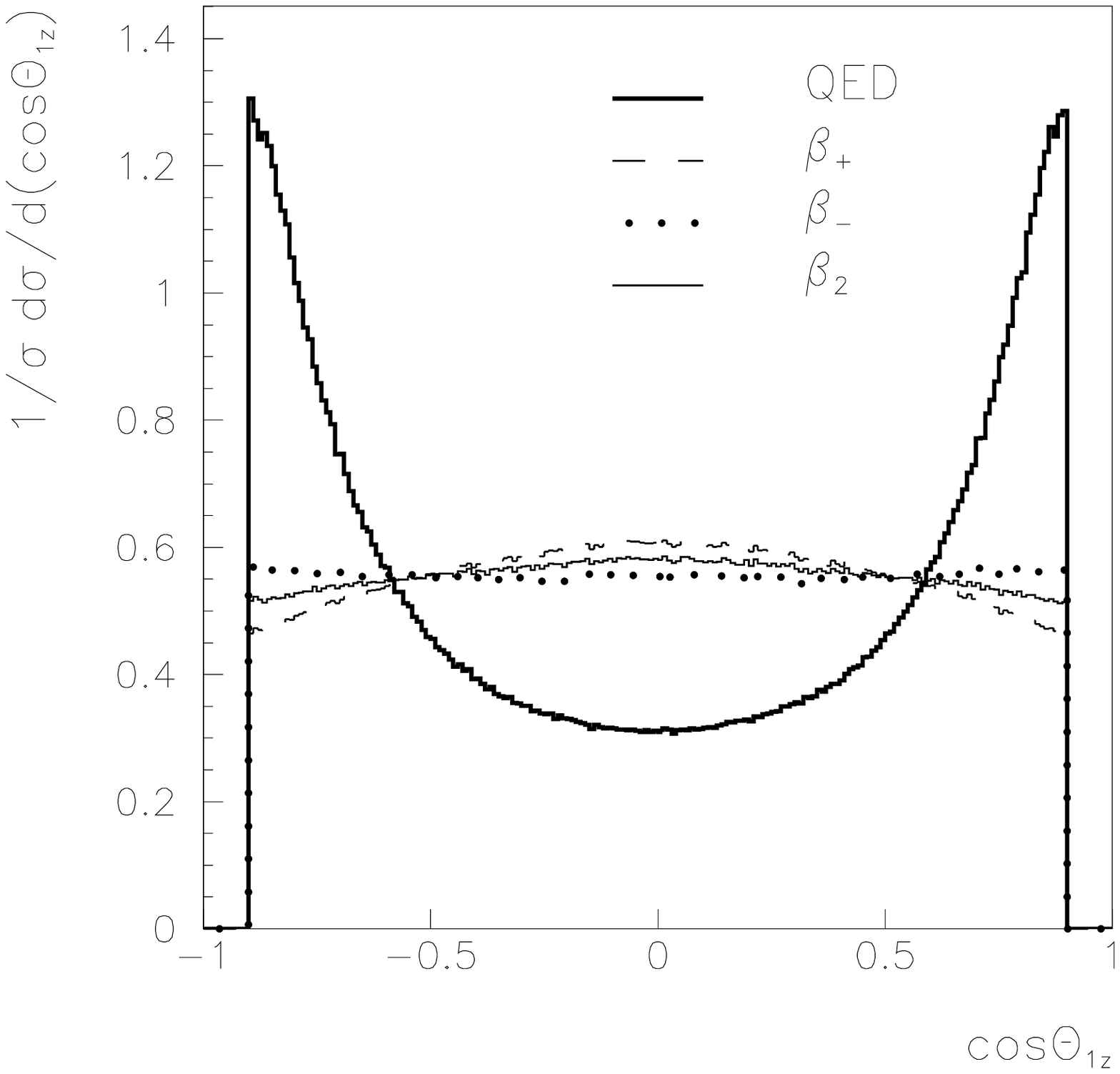}}
 \vspace*{0.5cm}
 \mbox{\epsfxsize=14cm\epsfysize=6.3cm\epsffile{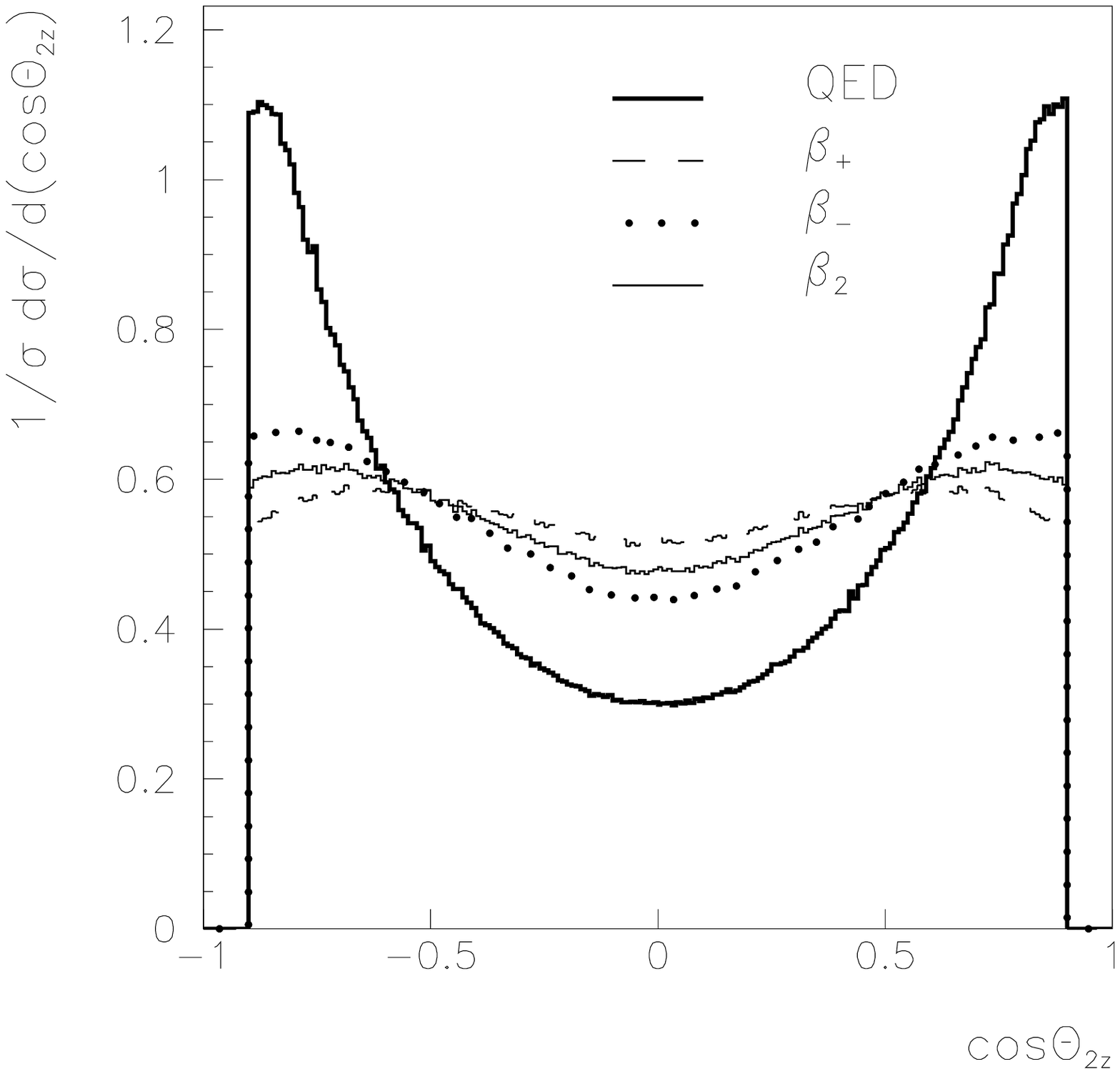}}
 \vspace*{0.5cm}
 \mbox{\epsfxsize=14cm\epsfysize=6.3cm\epsffile{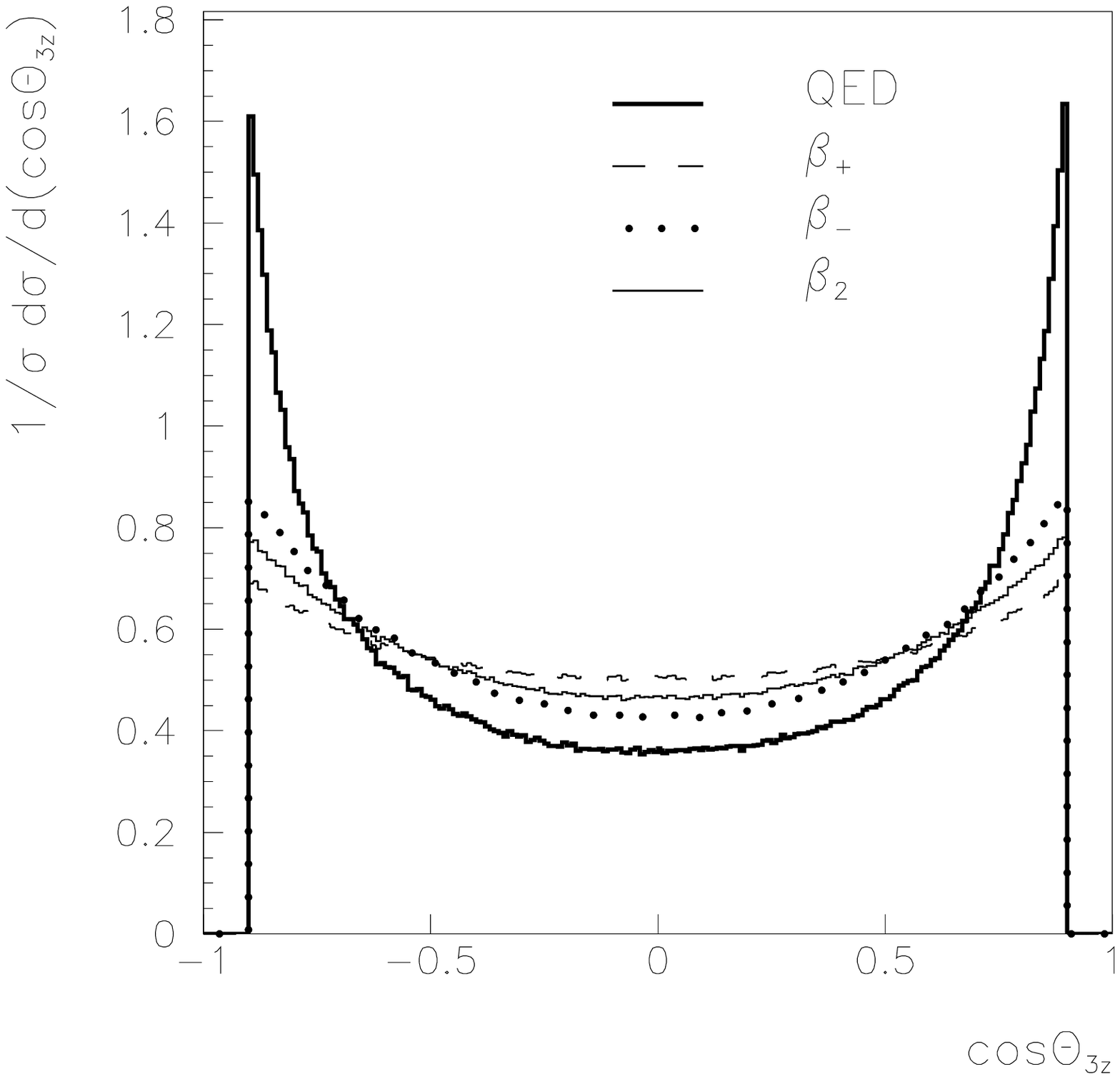}}
\end{center}
\end{figure*}
The QED events, as Figs.~\ref{disfbeam} show, are clearly peaked in the
forward-backward direction even after the cuts, whereas the anomalous events
are
rather central. Note should however be made that the softest non-QED photon has
also a slight tendency of preferring to be in the forward-backward region. In
this
respect it is really the most energetic photon distribution that shows the most
marked difference between the QED and the non-QED events. Once again, one sees
that all models for $Z$-initiated $3\gamma$ events give a sensibly similar
distribution and thus it is highly unlikely that one can resolve any difference
between the
models from the distributions that we have seen up to now. The same can be said
about the distribution in the acollinearity, once the cuts are applied.

\begin{figure*}[htbp]
\begin{center}
\caption{\label{disfcosn}{\em Distribution in the cosine of the angle 
between the beam and the normal to the event plane.}}
 \vspace*{0.5cm}
 \mbox{\epsfxsize=16.cm\epsfysize=15cm\epsffile{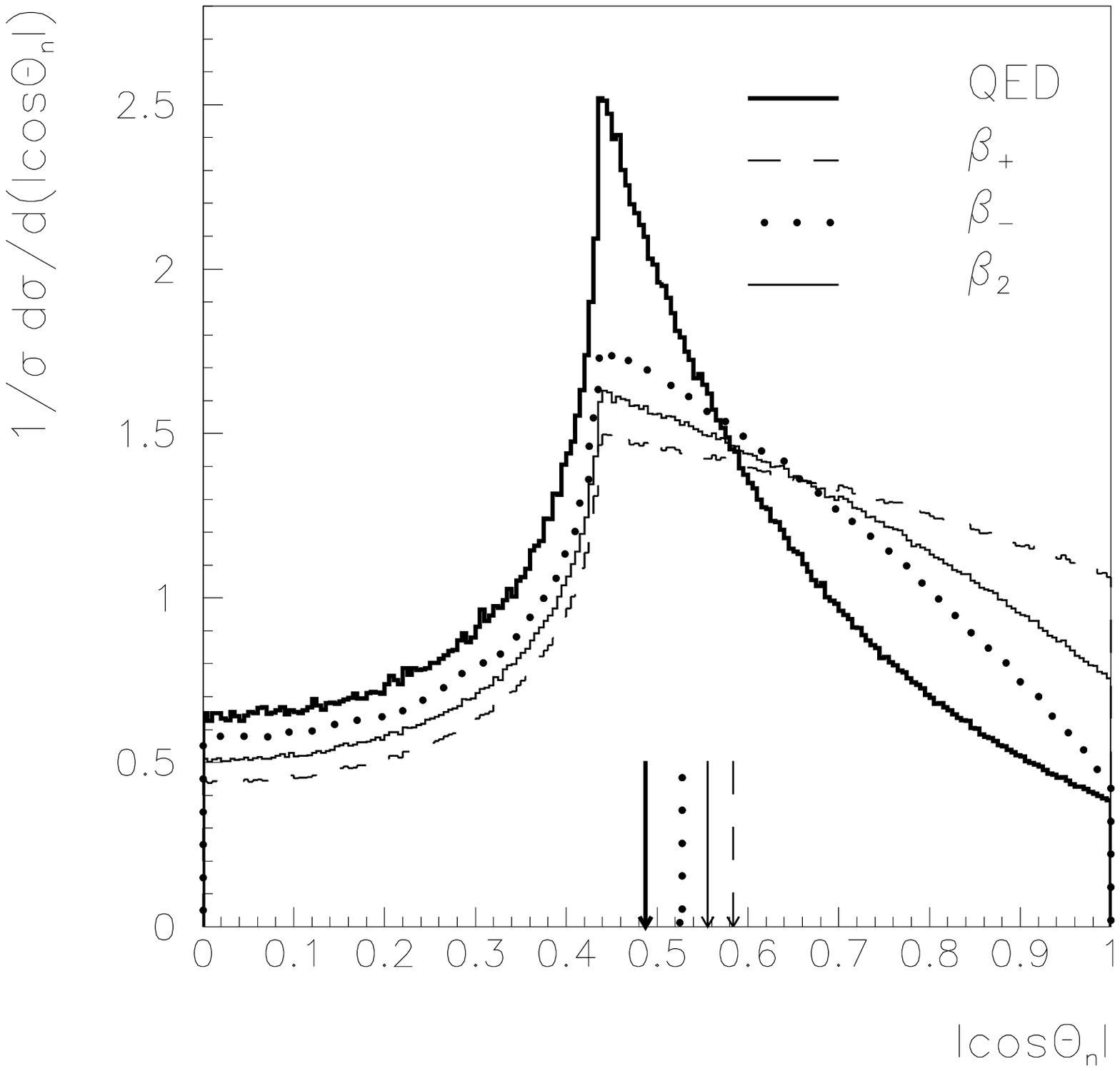}}
\end{center}
\end{figure*}

On  the other hand, the angular distribution of the event plane, or
equivalently the normal to
this plane, with respect to the beam axis, does show some interesting structure
that could be the best way of discriminating between the different models. This
is clearly seen in Fig.~\ref{disfcosn} , where we see that not only the average
values are
well separated but that there is a distinct lifting of the ``degeneracy"
between
the models, especially for values of the cosine of the angle between the plane
and the beam axis above 0.6.

\begin{figure*}[htbp]
\begin{center}
\caption{\label{scatter23}{\em
Normalised scatter plots that show the difference between the QED
distributions and those due to the independent $\beta_+$ and $\beta_-$
couplings.
The first is in the energy of the  softest photon ($E_3$) and the cosine
of the angle between the beam and the hardest photon ($\cos\theta_{1z}$). The
second set of scatter plots is in the energies of the softest ($E_3$) and
medium photon
($E_2$).}}
 \vspace*{0.5cm}
 \mbox{\epsfxsize=15.5cm\epsfysize=18cm\epsffile{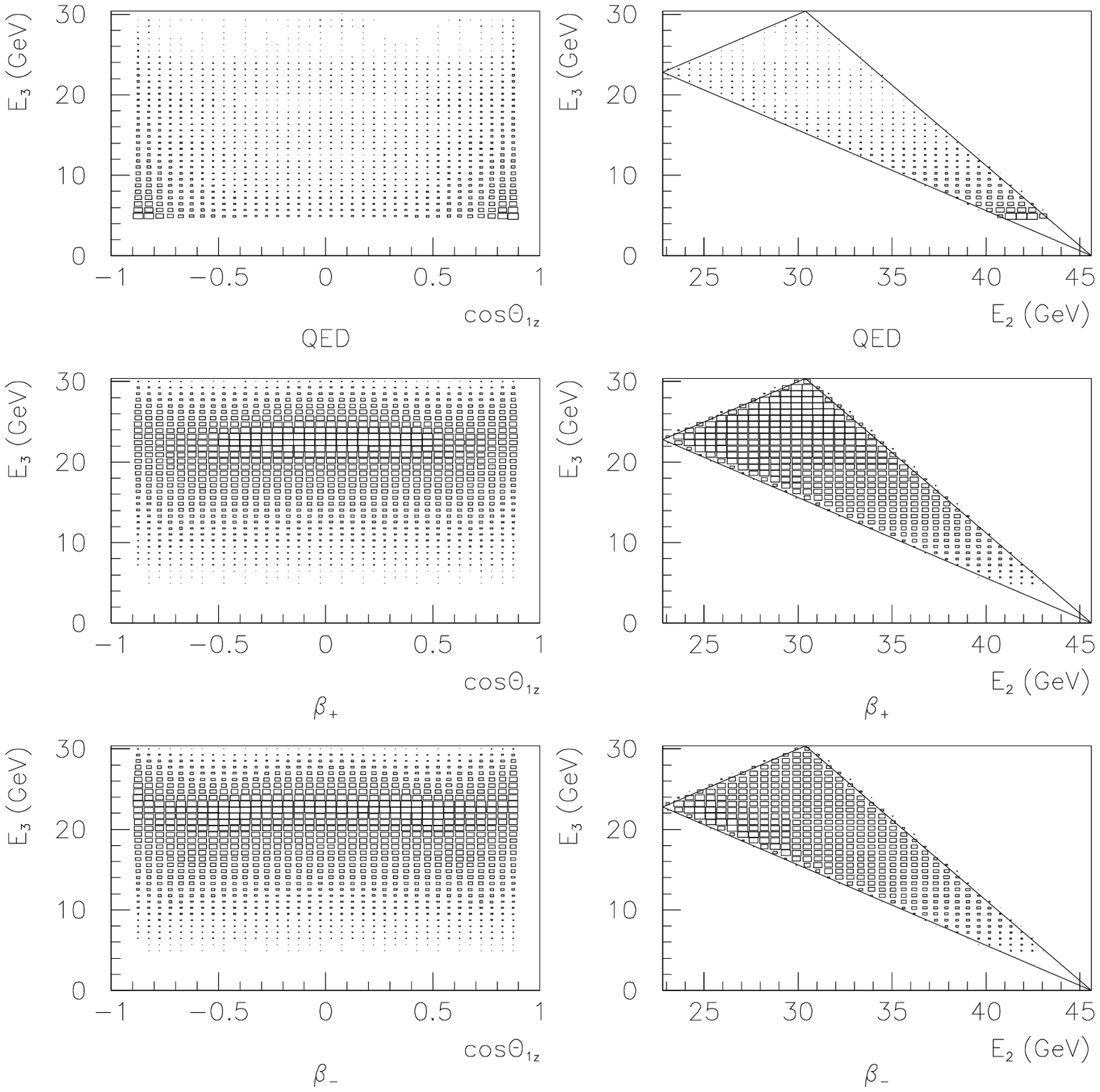}}
\end{center}
\end{figure*}

\begin{figure*}[htbp]
\begin{center}
\caption{\label{scatterne}{\em As in the previous scatter plots but as
variables: the cosine of the angle between the normal to the event plane and
the beam, and the energy of the softest or the hardest photon.}}
 \vspace*{0.5cm}
 \mbox{\epsfxsize=15.5cm\epsfysize=18cm\epsffile{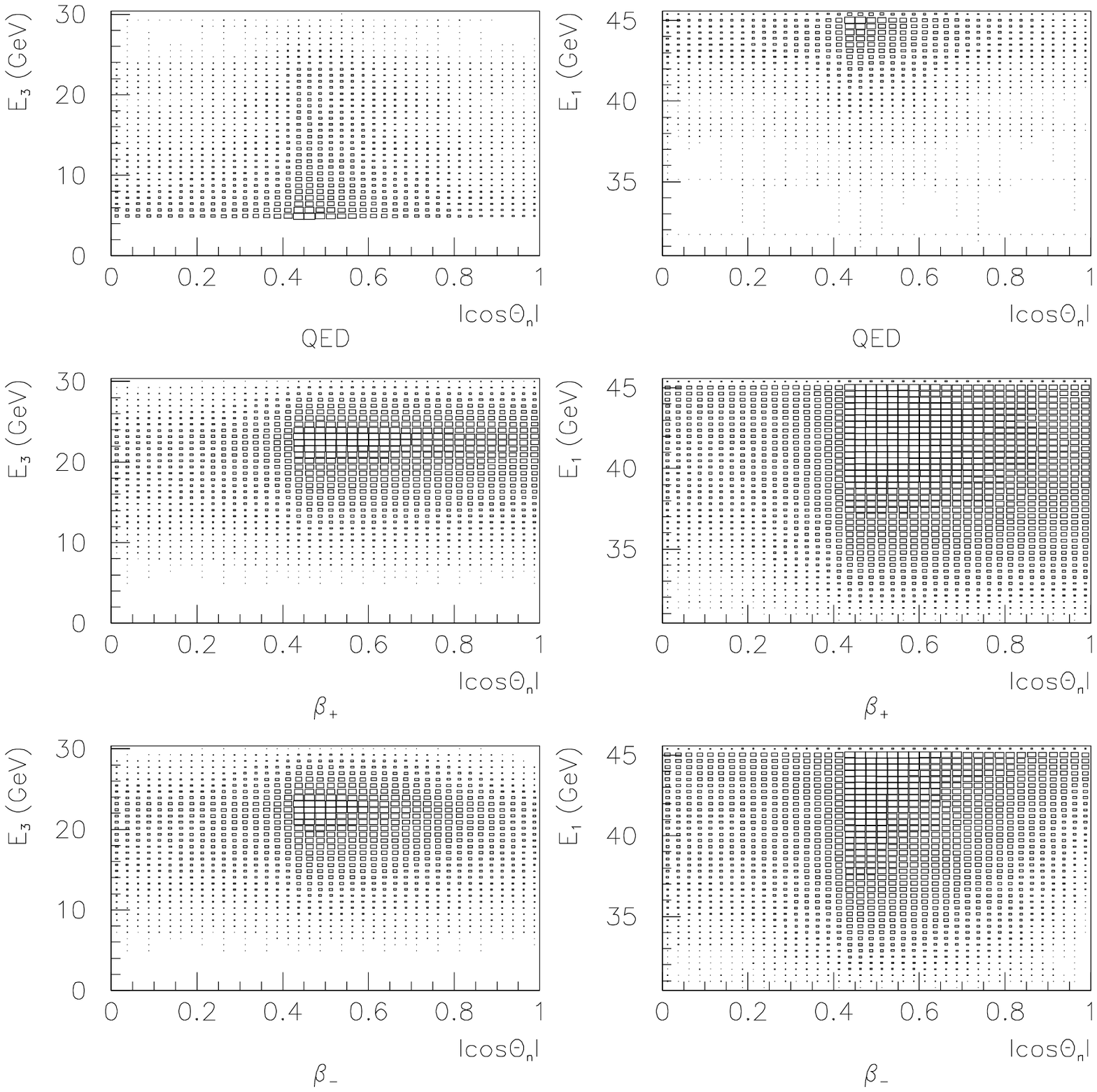}}
\end{center}
\end{figure*}

This lifting of the degeneracy with the devising of more optimal cuts to bring
out the
new physics and attempt to look at its origin can be made much clearer by
studying distributions in a combination of appropriate variables in the form of
scatter plots, for instance.
In the scatter plot of the energy of the two softest photons,
the pure QED and the $s$-channel $3\gamma$ are confined to two opposite corners
(see Fig.~\ref{scatter23}). Another obvious discrimination is to examine the
scatter plot involving the least energetic photon and the angle of the most
energetic photon with the beam.
The scatter plots involving the angle of the event plane with the beam show
what
could, in principle, be the best strategy to differentiate between various
models. We illustrate this in Fig.~\ref{scatterne} by taking as a second
variable
either the energy in the softest or the hardest photon.
\setcounter{subsection}{0}
\setcounter{equation}{0}
\def\thesubsection {\thesection.\arabic{subsection}}
\def\theequation{\thesection.\arabic{equation}}
\def\thefigure{\arabic{figure}}
\section{Conclusions and Discussion}
We have given a detailed analysis of the most general manifestation of New
Physics in the coupling of $Z \ra 3\gamma$. The analytical formulae
should help in quickly estimating the acceptances and providing a limit
on the operators from the ongoing LEP1 searches. We have
indicated how to optimise the searches (and the limits)
and in the eventuality of a signal how to disentangle various effects.
Our study also shows that for searches around LEP1 energies and with current
limits on the anomalous \z3gt couplings the effect of the interference between
New Physics and the irreducible QED $3\gamma$ background is abysmally
insignificant and can thus be neglected, although it is very straightforward
to include.

\begin{figure*}[htbp]
\begin{center}
\caption{\label{significances}{\em Statistical significances as a function of
the cms energy assuming a common effective luminosity $L=$100~pb$^{-1}$.
Three typical values of the branching ratio are assumed with
$\tilde{\beta}_+>0$.}}
 \vspace*{0.5cm}
 \mbox{\epsfxsize=15cm\epsfysize=9cm\epsffile{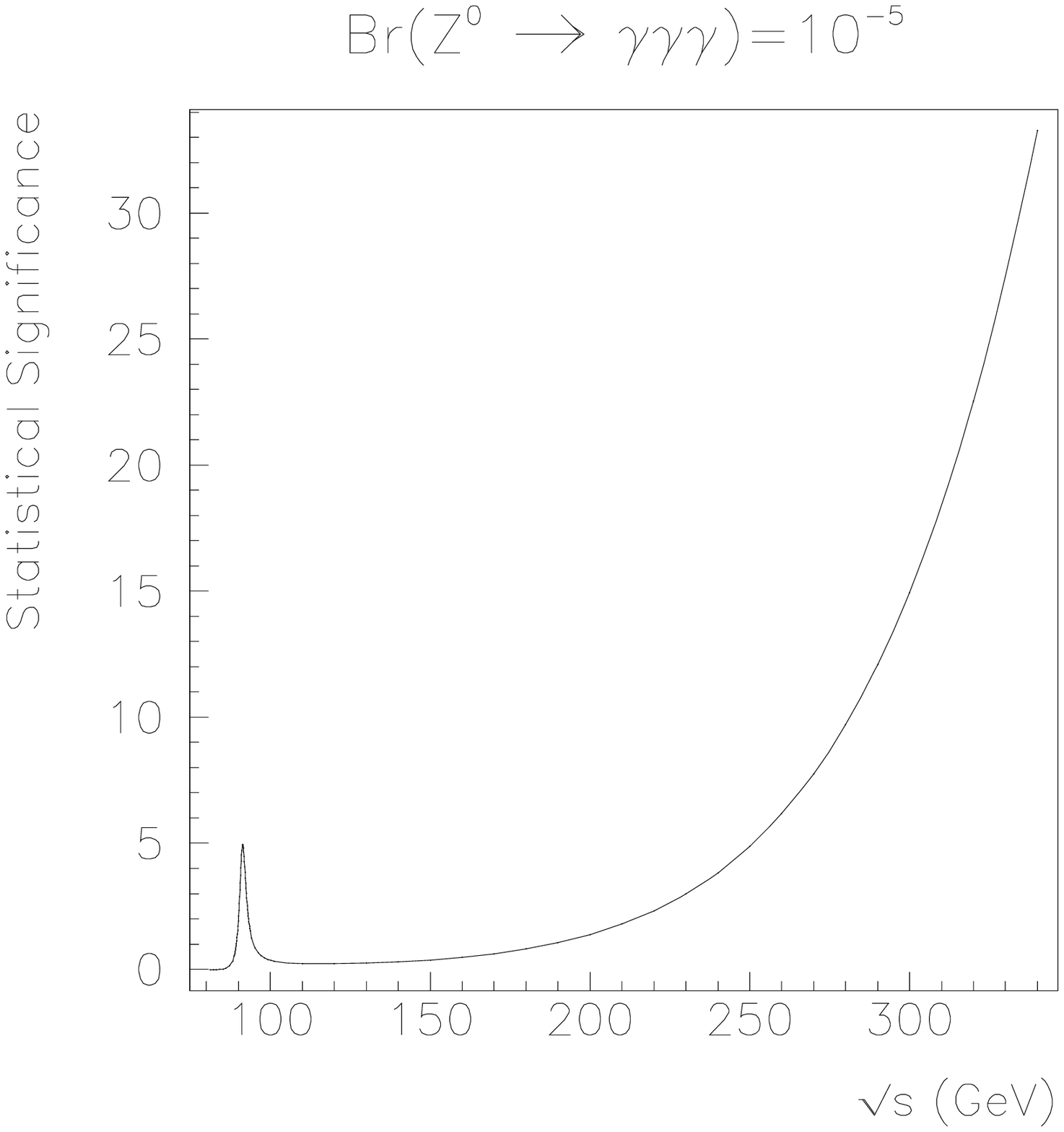}}
 \vspace*{0.5cm}
 \mbox{
  \mbox{\epsfxsize=7.5cm\epsfysize=9cm\epsffile{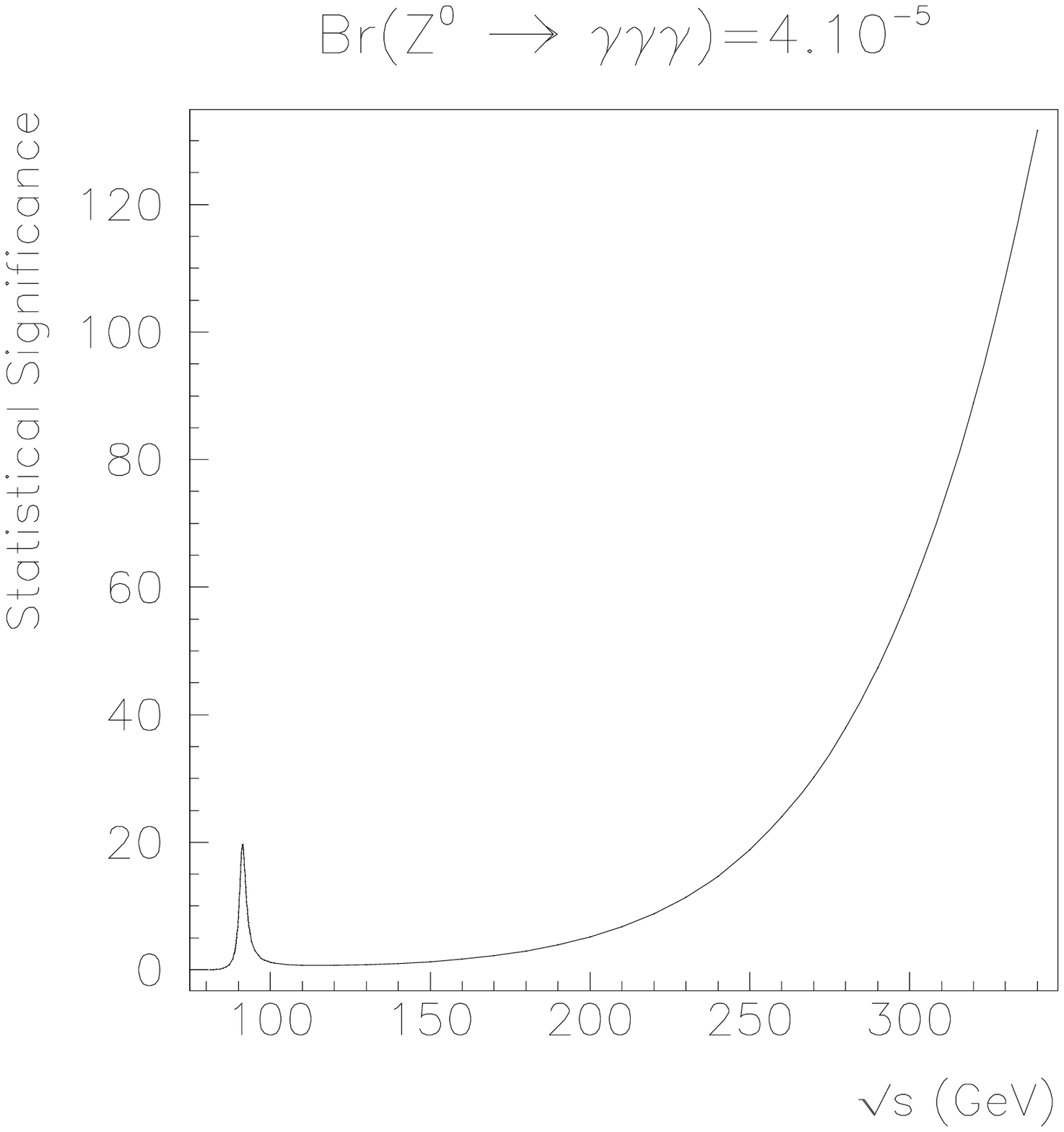}}
  \hspace*{0.5cm}
  \mbox{\epsfxsize=7.5cm\epsfysize=9cm\epsffile{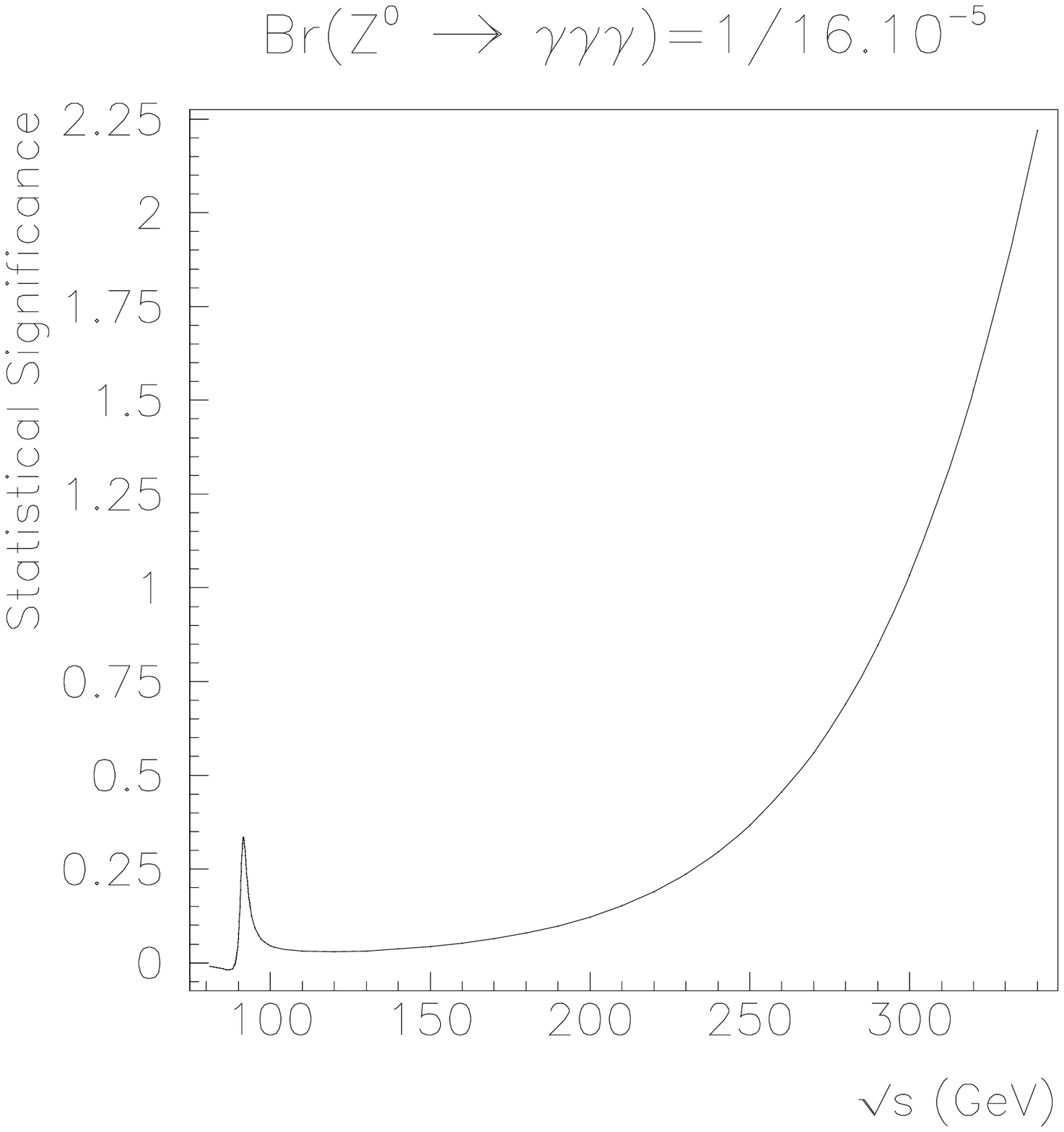}}}
\end{center}
\end{figure*}

Let us now ascertain the limits one would derive on the parameters of the
effective Lagrangian at the end of the LEP1 run. We will assume that each
experiment has collected a total luminosity $\int{\cal L}=$200~pb$^{-1}$ and
further assume a $100\%$ efficiency on the detection of the $3\gamma$ once the
experimental cuts defined by Eq.~\ref{expcuts} have been applied. Let us stress
that since all models, {\it i.e.}, both \bzpt$\;$ and \bzmt$\;$ give sensibly
the same
distributions and that since it will not be possible to reconstruct the
helicities of the energetic photons it is extremely difficult to distinguish
between the different $\tilde{\beta}$'s especially if the ``anomalous" signal
is
weak. Therefore, the limits will have to be extracted from the total cross
section after cuts have been applied. It ensues that  we will only have access
to the
combination of \bzpt$\;$ and \bzmt$\;$
$(\bzpc \sigma_+ + \bzmc \sigma_-)$
where $\sigma_\pm$ are defined in Eq.~\ref{ee3gsigs} and depend (mildly) on the
cuts
applied. Without the cuts one has  the combination that appears in the width,
$(3 \bzpc  + 5 \bzmc)$. With a healthy  statistics and a good signal it could
be
possible by judiciously varying the cuts to ``reconstruct" the correct
combination of \bzmt$\;$ and \bzpt$\;$. In what follows we will restrict the
discussion to one parameter.
Requiring a
$5\sigma$ deviation, with only the statistical error taken into account, one
finds for
$\tilde{\beta}_-$ for instance
\beq
\frac{4\alpha^2}{m^4} |\tilde{\beta}_-| < 0.215\;\;\;\;\; (0.256 \;\;\hbox{for}
\int{\cal
L}=100~\mbox{\rm pb}^{-1}).
\eeq
The increase in luminosity does not, unfortunately, tremendously improve the
limits, as the
effect of the New Physics is quadratic in the couplings and therefore the
sensitivity scales only as ${\cal L}^{1/4}$. This remains
true, as we have seen, even when one moves away from the $Z$ peak. Thus,  one
may wonder, since the QED cross section decreases with energy as $1/s$
while the anomalous grows as $s^3$, whether one could set a better limit at
LEP2. We have already answered by the negative. We show in
Fig.~\ref{significances} how the significances are changed as the energy is
increased while keeping the same cuts and normalising to a common integrated
luminosity of ${\cal L}=$100~pb$^{-1}$\footnote{Of course, as the energy
increases
one expects the luminosity to improve as well, however we stress again that
since the effect of the interference is marginal, the significance can be
easily
calculated through the scaling law ${\cal L}^{1/4}$.}. As the figures
demonstrate, for various values of $\beta_+$ (in the range set by LEP1), the
statistical significances are never better at LEP2 than what they are at LEP1.
One will take full advantage
of the energy increase of the anomalous cross section, bettering the
resonance enhancement, only for energies around $300$~GeV. Taking again the
operator defined by $\tilde{\beta}_-$, and with LEP2 meaning
$\sqrt{s}=190$~GeV, ${\cal L}=$500~pb$^{-1} $ a limit based on a $5\sigma$
(statistical
only) deviation gives
\beq
\frac{4\alpha^2}{m^4} |\tilde{\beta}_-|< 0.365,
\eeq
{\it i.e.}, worse than at LEP1. However, at a Next Linear Collider (NLC)
operating at 350~GeV with ${\cal L}=$10~fb$^{-1} $ we find that the LEP1 limit
is
improved by as much as an order of magnitude:
\beq
\frac{4\alpha^2}{m^4} |\tilde{\beta}_-| < 0.025.
\eeq
This said, it is not excluded that the strength of the 4-$\gamma$ vertex is
much
larger than that of the \z3gt vertex, in which case $3\gamma$ production,
proceeding through photon exchange, will be much better studied than at LEP1,
since it does not receive any resonant enhancement.
Therefore we should urge the LEP2 collaborations to scrutinise $3\gamma$
production. The topology of these $3\gamma$ events is {\em exactly} the
same as those originating from the $\epem \ra Z\ra 3\gamma$ that we have
studied
here; in particular, all the normalised distributions are the same. To adapt
what
we have investigated in this paper to the effect of a 4-$\gamma$ vertex
in 3-$\gamma$ production
one simply makes the replacement $\tilde{\beta}_\pm\ra \beta_\pm$ together with
the
change of the $Zee$ vertex into the electromagnetic one and the
replacement of the $Z$ propagator.
For instance assuming that $\tilde{\beta}_\pm=s_W/c_W \beta_\pm$ as given by a
model with a strongly
interacting SU(2) singlet fermion, then the 4-$\gamma$ strength is larger than
that of the
\z3gt. For this particular case the combined effect of the photon and the $Z$
is accounted for by the factor
\beqn
\frac{g_L^2 \left(1-\displaystyle\frac{c_W^2}{g_L}
(1-\displaystyle\frac{M_Z^2}{\displaystyle s})\right)^2
+g_R^2 \left(1-\displaystyle\frac{c_W^2}{\displaystyle s_W^2}
 (1-\displaystyle\frac{M_Z^2}{\displaystyle s})\right)^2}
{g_L^2+g_R^2}
\eeqn
that multiplies the results of taking into account the $Z$-channel only.
At 190~GeV this  enhances the $Z$ cross section by almost
seven fold.
At LEP2, there is also the possibility of
checking for the 4-$\gamma$ vertex through $2\gamma$ production in the
(peripheral) $\gamma \gamma$ processes. A dedicated investigation of this
coupling for LEP2 and beyond is under investigation  \cite{NousGinzburg4g}.

\noindent {\large \bf Acknowledgement:}
\noindent We thank Genevi\`eve B\'elanger for contributing to some parts of
this
paper. We gratefully acknowledge the stimulating discussions we had with her
and
with Ilya Ginzburg. We also thank Peter M\"attig for his careful reading of the
manuscript and for his comments.
This paper would not have seen the ``light" were it not for the
continuous insistence of the many experimentalists within Aleph, Delphi, L3 and
Opal
who urged us to provide this detailed analysis and for suggesting to adapt our
results into the form of a Monte Carlo. We would particularly like to thank,
for
their suggestions and patience, F.~Bar\~{a}o, P.~Checchia, B.~Hartman,
K.~Kawagoe,
Y.~Kariokatis, R.~Rosmalen and G.~Wilson.

\setcounter{section}{0}
\setcounter{subsection}{0}
\setcounter{equation}{0}
\def\thesection{\Alph{section}}
\def\thesubsection {\thesection.\arabic{subsection}}
\def\theequation{\thesection.\arabic{equation}}

\newpage
\noindent {\Large \bf Appendix:}

\section{Helicity amplitudes with the spinor inner product technique.}

With the cuts that we had to implement, the helicity amplitudes could be
evaluated by neglecting the electron mass. There is a variety of methods
for the evaluation of the helicity amplitudes
\cite{GastmansWu,Manganophysrep}.
The most powerful are
those that have been developed for the calculation of multiparticle processes
especially in massless QCD  \cite{Kleiss,XZC,Manganophysrep} and are thus well
adapted for the process at hand. The amplitudes are written in terms of
Spinor Inner Products. We have performed all the helicity amplitudes
calculation
with the help of FORM. We have also checked our results against known results
for the QED part and against the usual trace technique summation for the
matrix element squared. For the helicity technique we have made full use of
gauge invariance by choosing the most appropriate choice of gauge for the
polarisation vectors to render the
expressions as simple as possible. This is a huge gain on the usual squaring
technique.
The helicity amplitudes can all be expressed though the  inner product
$S(p,q)$, where $p,q$ are (light-like) momenta. Following the handy notation of
  \cite{XZC}, the massless spinors are written as
\beqn
|p\pm>\equiv u\left( p,\pm \frac{1}{2})=v(p,\mp \frac{1}{2} \right) \nonumber
\\
<p \pm|\equiv \bar{u}\left( p,\pm \frac{1}{2})=\bar{v}(p,\mp \frac{1}{2}
\right)
\eeqn
then
\beqn
S(p,q)&=&<p-|\,q+>=<q+|\,p->^*=-S(q,p)\;\;\;;\;\;\; \nonumber \\
 |S(q,p)|^2&=&2 p\cdot q \;\;\;;\;\;\; (p^2=q^2=0).
\eeqn
$S$ is antisymmetric, unlike in   \cite{Kleiss}.

We decompose the amplitudes into the standard QED part and the two
non-interfering $\tilde{\beta}_\pm$ as
\beqn
{\cal A}=-\left(8 \pi \alpha \right)^{3/2} \left\{ {\cal A}^{QED}
\;-\;\frac{\alpha}{\pi M^4}\sqrt{\frac{\alpha(s)}{\alpha}}\,
\frac{1}{s_W c_W}
 D_Z(s) \left( \tilde{\beta}_+{\cal B}^+
+\tilde{\beta}_-{\cal B}^-\right) \right\}.
\eeqn

The helicity components are labelled as
${\cal A}_{\sigma_1,\sigma_2;\lambda_1,\lambda_2,\lambda_3}$, where the first
labels  refer to the helicity of the $e^-$ and $e^+$.
Chirality ($m_e=0$) only allows the helicity amplitudes
${\cal A}_{\sigma,-\sigma;\lambda_1,\lambda_2,\lambda_3}$. One then only needs
the
expressions for ${\cal A}_{+,-; \pm,\pm,\pm},{\cal A}_{+,-; +,-,-}$ and
${\cal A}_{+,-;++-}$.
For QED we recover the known result that the amplitude with the like-sign
photon
helicities vanishes:
\beqn
{\cal A}_{+,-; \pm,\pm,\pm}^{QED}=0
\eeqn
while the two others are simply given by
\beqn
\label{helqed}
{\cal A}^{QED}_{+-;--+}&=&-
\frac{S(p_1,p_2)S(p_1,k_3)^2}{S(p_1,k_1)S(p_1,k_2)S(p_2,k_1)S(p_2,k_2)}
\nonumber \\
{\cal A}^{QED}_{+-;++-}&=&\left(
\frac{S(p_1,p_2)S(p_2,k_3)^2}{S(p_1,k_1)S(p_1,k_2)S(p_2,k_1)S(p_2,k_2)}
\right)^*
\eeqn

For the anomalous part, ${\cal B}^-$ only contributes to the like-sign
photon helicity amplitude whereas ${\cal B}^+$ only for mixed helicity states:
\beqn
\label{samehelano}
{\cal B}^+_{+-;+++}={\cal B}^+_{+-;---}=
  {\cal B}^-_{+-;++-}= {\cal B}^-_{+-;+--}=0
\eeqn
and
\beqn
{\cal B}^+_{+-;--+}&=&g_R\;\;
 S(p_1,p_2)^* (S(k_3,p_1) S(k_1,k_2)^*)^2 \nonumber \\
{\cal B}^+_{+-;++-}&=&-g_R\;\;
S(p_1,p_2) (S(k_3,p_2)^* S(k_1,k_2))^2
\eeqn
\beqn
{\cal B}^-_{+-;+++}&=&g_R \;\;S(p_1,p_2)^*\;
\left(
 S(k_2,k_3)^2 S(k_1,p_1)^2 \right. \nonumber \\
& & \;\;\;\;\;\;\;\;\ \left. +S(k_1,k_3)^2 S(k_2,p_1)^2
+S(k_1,k_2)^2 S(k_3,p_1)^2 \right) \nonumber \\
{\cal B}^-_{+-;---}&=&-g_R \;\;S(p_1,p_2)\;
\left (
 S(k_2,k_3)^2 S(k_1,p_2)^2 \right. \nonumber \\
& & \;\;\;\;\;\;\;\;\ \left. +S(k_1,k_3)^2 S(k_2,p_2)^2
+S(k_1,k_2)^2 S(k_3,p_2)^2 \right )^*. \nonumber \\
\eeqn

All other amplitudes are found by permutation in the photons and parity
conjugation that reverses all signs of the helicity. For the $Z$ mediated
amplitudes
this conjugation should also be accompanied by allowing for $g_R \ra g_L$.

The numerical  generation of $S(p,q)$ is done through the definition
\beqn
S(p,q)=\frac {(p_1+ip_2)} {\sqrt{(p_0+p_3)}} \sqrt{(q_0+q_3)}
  -\frac {(q_1+iq_2)} {\sqrt{(q_0+q_3)}} \sqrt{(p_0+p_3)}\;\;\;;\;\;\;
p=(p_0,p_1,p_2,p_3).
\eeqn

\vspace*{0.5cm}
\newpage

\noindent {\large \bf Where to find the generator?}

\noindent
All the cross sections and distributions found in this paper have been
obtained by numerical integration of  helicity-amplitude-based matrix
elements.  This was necessary in order to show that the interference
between the QED background and the anomalous signal is negligible
for energies up to LEP2  and for reasonable values of the couplings.
However, from an experimentalist point of view, this is not exactly
what is needed.  To conduct a detailed experimental analysis, it is more
important
to have an event generator for the QED background and
a second one for the signal.  The former can already be
found on the market\cite{generatorBK} and we now provide a new generator for
the anomalous
signal.  It can be found at
lapphp0.in2p3.fr/pub/preprints-theorie/ee3gammagenerator.uu using
anonymous ftp.  The uu-encoded file contains five files of which only
one is important.  The four others are there only to run a small
demonstration program to see if it runs well on your system.

\end{document}